\documentclass[%
preprint,
 amsmath,amssymb,
 aps,
]{revtex4-1}

\usepackage{graphicx}
\usepackage{dcolumn}
\usepackage{bm}

\usepackage{amsmath}
\usepackage{amssymb}
\usepackage{mathrsfs}

\makeatletter
 \renewcommand{\theequation}{\arabic{section}.\arabic{equation}}
 \@addtoreset{equation}{section}
 
\makeatother
\newcommand{\yna}{{}^{\scalebox{0.5}{Y}}\nabla}
\newcommand{\bas}{{}^{\scalebox{0.5}{B}}}
\newcommand{\sol}{{}^{\scalebox{0.5}{S}}}

\newcommand{\mathsfbi}[1]{\boldsymbol{\mathsf{#1}}}
\newcommand{\vc}[1]{\boldsymbol{#1}}
\newcommand{\cnv}[3]{\left[\vc{#1};\vc{#2},\vc{#3}\right]}
\newcommand{\bepsilon}{\bas\!\varepsilon}
\newcommand{\bask}{\bas\!\!\mathscr{K}}
\newcommand{\norm}[1]{\lVert #1 \rVert}
\newcommand{\TS}{\overset{{}^{\textrm{TS}}}{\longmapsto}}
\newcommand{\mlp}{y}
\newcommand{\mlpind}{{}^{\scalebox{0.5}{$y$}}}
\begin{document}


\title{Constitutive theory of inhomogeneous turbulent flow\\based on two-scale Lagrangian formalism}

\author{Taketo ARIKI}
 \email{ariki@cfd.mech.tohoku.ac.jp} 
 \affiliation{Institute of Materials and Systems for Sustainability,\\
 Nagoya University, Furo-cho, Chikusa-ku, Nagoya, Japan}
\date{\today}

\begin{abstract}
A self-consistent closure theory is developed for inhomogeneous turbulent flow, which enables systematic derivations of the turbulence constitutive relations without relying on any empirical parameters. The double Lagrangian approach based on the mean and fluctuation velocities allows us to describe a wide variety of correlations in a consistent manner with both Kolmogorov's inertial-range scaling and general-covariance principle.

\end{abstract}
\maketitle

\section{Introduction}
Turbulent flow represents highly disordered behaviors in both space and time. Whereas believed --- not yet proven --- to be deterministic solution of the Navier-Stokes equation, understanding and control of turbulence often require its statistical treatment, where the \emph{closure problem} is unavoidable in constructing statistical governing laws.

A number of closure theories have been developed for homogeneous turbulence, where one can analyze small-scale universality without interference from non-uniform feature in large-scale properties. Among others, direct interaction approximation (DIA) gave a breakthrough by incorporating stochastic-relaxation process using infinitesimal response function \cite{Kraichnan59}. Reconstruction of DIA from field-theoretic approach \cite{Wyld61} and Lagrangian description gave rise to the most successful class of moment-closure theories (e.g. Lagrangian-history DIA (LHDIA) \cite{Kraichnan65,Kraichnan66}, strain-based LHDIA \cite{KH78}, and Lagrangian renormalized approximation (LRA) \cite{Kaneda81,Kaneda86}) which succeeded in deriving Kolmogorov's energy spectrum \cite{K41a} with its proportional constant (Kolmogorov constant) in a self-consistent manner. Another theoretical development has arisen from renormalization group (RNG) theory which elucidates relevant physical properties of scale-similar behaviors \cite{FNS76}. Further extension of RNG was performed to determine universal constants \cite{YO86prl,YO86jsc}, which although suffers from an inconsistency with Kolmogorov's timescale due to its Eulerian formulation \cite{Eyink94}. 


Unlike well-established theories of homogeneous turbulence, most of inhomogeneous-turbulence closures are based on heuristic approaches, while there are some analytical attempts based on exact governing laws; direct application of DIA to shear- and thermally-driven turbulence \cite{Kraichnan64a}, coordinate-decomposed DIA applied to channel turbulence \cite{Leslie73}, etc. 
Two-scale DIA (TSDIA), a combination technique of multiple-scale expansion and DIA, offers more feasible approach to inhomogeneous turbulence via one-point closure method, resulting in wider applicability to general turbulence accompanied by complex flow geometries \cite{Yoshizawa84,Hamba87}. 



TSDIA provides a systematic closure scheme for arbitrary unclosed correlation in terms of second-order moments. Using deductive approach based on the exact governing laws, TSDIA has uncovered non-trivial physics inaccessible from phenomenological understandings: non-linear turbulence viscosity effect \cite{Yoshizawa84,NY87,Okamoto94}, counter-gradient diffusion of passive scalar \cite{Yoshizawa85,Hamba93},  turbulence viscosity near shock wave \cite{Yoshizawa90}, vortex dynamo effect \cite{YY93}, and cross-helicity dynamo effect in MHD turbulence \cite{YY93m,YB16}. Particularly, for charge-neutral incompressible fluid, TSDIA derives $\mathscr{K}$-$\varepsilon$ type models with their proportional constants; e.g. applying TSDIA to the Reynolds stress $\vc{R}(\equiv\langle \vc{v}'\otimes \vc{v}'\rangle)$ yields the following series expansion:
\begin{equation}
\begin{split}
&R_{ij}=\frac{2}{3}\mathscr{K}\delta_{ij}-0.123 \frac{\mathscr{K}^2}{\varepsilon}
\left(\frac{\partial V_i}{\partial x_j}+\frac{\partial V_j}{\partial x_i}\right)\\
&+0.0427\frac{\mathscr{K}^3}{\varepsilon^2}\frac{D}{Dt}
\left(\frac{\partial V_i}{\partial x_j}+\frac{\partial V_j}{\partial x_i}\right)\\
&+0.0542\frac{\mathscr{K}^3}{\varepsilon^2}
\left(\frac{\partial V_i}{\partial x_a}\frac{\partial V_j}{\partial x_a}-\frac{1}{3}\delta_{ij}\frac{\partial V_a}{\partial x_b}\frac{\partial V_a}{\partial x_b}\right)\\
&+0.0297\frac{\mathscr{K}^3}{\varepsilon^2}
\left(\frac{\partial V_i}{\partial x_a}\frac{\partial V_a}{\partial x_j}+\frac{\partial V_j}{\partial x_a}\frac{\partial V_a}{\partial x_i}-\frac{2}{3}\delta_{ij}\frac{\partial V_a}{\partial x_b}\frac{\partial V_b}{\partial x_a}\right)\\
&+\cdots,
\end{split}
\label{TSDIA}
\end{equation}
where $\vc{V}(\equiv\langle\vc{v}\rangle)$ is the mean velocity, $\mathscr{K}(\equiv \langle \norm{\vc{v}'}^2 \rangle/2)$ is turbulence energy, $\varepsilon(\equiv \nu \langle \norm{\vc{\nabla}\otimes\vc{v}'}^2\rangle)$ is its dissipation rate. Here all the numerical constants are analytically derived. Likewise, TSDIA enables analytical closure of the turbulence constitutive relation, which may be appreciated as its remarkable advantage over traditional modeling strategies based on dimensional and tensor analyses. 

In spite of these prominent successes, TSDIA formalism, as a theory of physics, suffers from critical caveats to be carefully improved. In this paper, we focus on the following two aspects above all else:
\begin{enumerate}
\item TSDIA is inconsistent with some classes of coordinate transformations. At least it has been pointed out so far that TSDIA results do not transform in a correct way under the time-dependent rotation \cite{HS08}. More general criticism may be made from the view point of the \emph{covariance principle} \cite{Frewer09,Frewer16,Ariki15a}. It is generally stated that any class of turbulence-constitutive model should satisfy the covariance principle so that the mean flow calculated by the model is consistent with the physical objectivity \cite{Ariki15a}. However, TSDIA contradicts the covariance principle of turbulence, and its physical prediction depends on the coordinate frame.

\item TSDIA has its roots in the Eulerian DIA inconsistent with inertial-range scaling of Kolmogorov's theory \cite{K41a} based on Lagrangian picture. To avoid the fatal contradiction in the scaling law, TSDIA relies on an artificial removal of the infrared divergence caused by the sweeping effect \cite{Yoshizawa78,Yoshizawa84}.
\end{enumerate}
Note that these two shortfalls arise from a common fact; TSDIA is fully based on the Eulerian framework. Then what is needed for further progress is to change the coordinate representation. The key ingredient is the double-Lagrangian formalism proposed by Ref. \cite{Ariki17} where two independent Lagrangian approaches are introduced on the basis of instantaneous and mean flows (see Sec. VI B and VI C of Ref. \cite{Ariki17}).  Mean-Lagrangian picture based on the mean flow, on one hand, guarantees covariant description of the history effect on the mean flow in an analogous way to general continuum physics. Fine-Lagrangian picture corresponding to the instantaneous flow, on the other hand, cancels out the known sweeping effect, removing contradiction with the Kolmogorov theory.

In this paper, we develop a turbulence-constitutive theory consistent with both covariance principle and the Kolmogorov theory with the help of the double-Lagrangian formalism. The resultant theory is a combination technique of two-scale expansion and LRA -- hereafter referred to as \emph{TSLRA} -- providing an overall reconstruction of TSDIA theory with fully keeping its advantages. In Sec. \ref{FORMULATION}, we provide a general formulation of TSLRA, which is applied to, as one of the most interesting examples, the Reynolds stress in Sec. \ref{REYNOLDS STRESS}.


\section{Formulation}\label{FORMULATION}

In this section, TSLRA will be formulated following the several steps; (i) rewriting the fluctuation equations in the mean-Lagrangian coordinate system (\ref{GOVERNING EQUATIONS}-\ref{MEAN LAGRANGIAN}), (ii) applying the two-scale representation to the fluctuation equations (\ref{TWO-SCALE REPRESENTATION}), (iii) rewriting the two-scale equations into a perturbed Navier-Stokes equation (\ref{FOURIER TRANSFORM}-\ref{ORTHONORMAL}), and (iv) constructing a renormalized perturbation theory on the basis of the perturbed Navier-Stokes equation (\ref{PERTURBATION}-\ref{RENORMALIZATION}). The obtained renormalized perturbation theory expresses turbulence correlations in terms of the second-order statistics of the velocity fluctuation.

\subsection{Governing equations}\label{GOVERNING EQUATIONS}
In this paper we apply the Reynolds decomposition to field variables. Given the velocity field $\vc{v}$ and the pressure $\mathrm{p}$ as dynamical variables, these are decomposed as
\begin{equation*}
\vc{v}=\left(\vc{v}-\vc{V}\right)+\vc{V}=\vc{v}'+\vc{V},\ \ 
\mathrm{p}=\left(\mathrm{p}-\mathrm{P}\right)+\mathrm{P}=\mathrm{p}'+\mathrm{P},
\end{equation*}
where $\vc{V}\,(\equiv\langle\vc{v}\rangle)$ and $\mathrm{P}\,(\equiv\langle\mathrm{p}\rangle)$ are the mean velocity and pressure defined by ensemble averaging $\langle \cdots \rangle$, $\vc{v}'\,(\equiv\vc{v}-\vc{V})$ and $\mathrm{p}'\,(\equiv\mathrm{p}-\mathrm{P})$ are the velocity and pressure fluctuations. As proven by Ref. \cite{Ariki15a}, velocity fluctuation behaves as a \emph{true} vector field under coordinate transformations maximally allowed in the non-relativistic regime, and its governing law can be rewritten in generally covariant form. Let $\{\vc{x}\}(\equiv\{x^{\scalebox{0.6}{1}},x^{\scalebox{0.6}{2}},x^{\scalebox{0.6}{3}}\})$ be a set of general coordinates. The velocity and pressure fluctuations obey the following set of equations \cite{Ariki15a}:
\begin{equation}
\begin{split}
&\left( \frac{\mathfrak{D}}{\mathfrak{D} t}-\nu\Delta\right)v'^i(\vc{x},t)+\left(v'^i v'^j \right)_{;j}(\vc{x},t)+\mathrm{p}'^{,i}(\vc{x},t)\\
&=-\left(S^i_j+\Theta^i{}_j\right)(\vc{x},t)\ v'^j(\vc{x},t)+R^{ij}{}_{;j}(\vc{x},t),
\end{split}
\label{v' eq.}
\end{equation}
\begin{equation}
v'^i{}_{;i}(\vc{x},t)=0,
\label{incomp}
\end{equation}
Where commas (,) and semicolons (;) in the index notation express partial and covariant derivatives; the covariant derivative operates on an arbitrary tensor field $T^{ij\cdots}{}_{mn\cdots}$ as
\begin{equation*}
\begin{split}
T^{ij\cdots}{}_{mn\cdots;a}(\equiv\nabla_a T^{ij\cdots}{}_{mn\cdots})
=T^{ij\cdots}{}_{mn\cdots,a}
&+\Gamma^i_{ab}T^{bj\cdots}{}_{mn\cdots}+\Gamma^i_{ab}T^{bj\cdots}{}_{mn\cdots}-\cdots\\
&-\Gamma^b_{am}T^{ij\cdots}{}_{bn\cdots}-\Gamma^b_{an}T^{ij\cdots}{}_{mb\cdots}-\cdots,
\end{split}
\end{equation*} 
where $\Gamma^i_{ab}=g^{ij}\Gamma_{j\cdot ab}$ is the Christoffel symbol of the Levi-Civita connection: $\Gamma_{j\cdot ab}=(g_{ja,b}+g_{jb,a}-g_{ab,j})/2$. A time-derivative operator $\mathfrak{D}/\mathfrak{D} t$ is the convective derivative based on the mean flow \cite{Ariki15a}:
\begin{equation*}
\begin{split}
\frac{\mathfrak{D}}{\mathfrak{D}t}T^{ij\cdots}{}_{mn\cdots}
=\left(\frac{\partial}{\partial t}+V^j\nabla_j\right)  T^{ij\cdots}{}_{mn\cdots}
&-V^i{}_{;a}T^{aj\cdots}{}_{mn\cdots}-V^j{}_{;a}T^{ia\cdots}{}_{mn\cdots}-\cdots\\
&+V^b{}_{;m}T^{ij\cdots}{}_{bn\cdots}+V^b{}_{;n}T^{ij\cdots}{}_{mb\cdots}+\cdots.
\end{split}
\end{equation*}
The mean strain rate $\mathsfbi{S}$ and the absolute vorticity $\boldsymbol{\Theta}$ are given respectively by $S_{ij}\equiv \mathfrak{D} g_{ij}/\mathfrak{D}t$ and $\Theta_{ij}\equiv z^I{}_{,i}\,z^J{}_{,j}(V_{I;J}-V_{J;I})$ ($V_I$: mean-velocity in an inertial frame of reference $\{\mathbf{z}\}$), respectively \cite{Ariki15a}. All derivative operations and quantities in Eqs. (\ref{v' eq.}) and (\ref{incomp}) are generally covariant, i.e. Eqs. (\ref{v' eq.}) and (\ref{incomp}) does not change its tensorial form under most general coordinate transformation within the non-relativistic limit.

\subsection{Mean-Lagrangian formalism}\label{MEAN LAGRANGIAN}
\begin{figure}
\centering
\includegraphics[width=12cm]{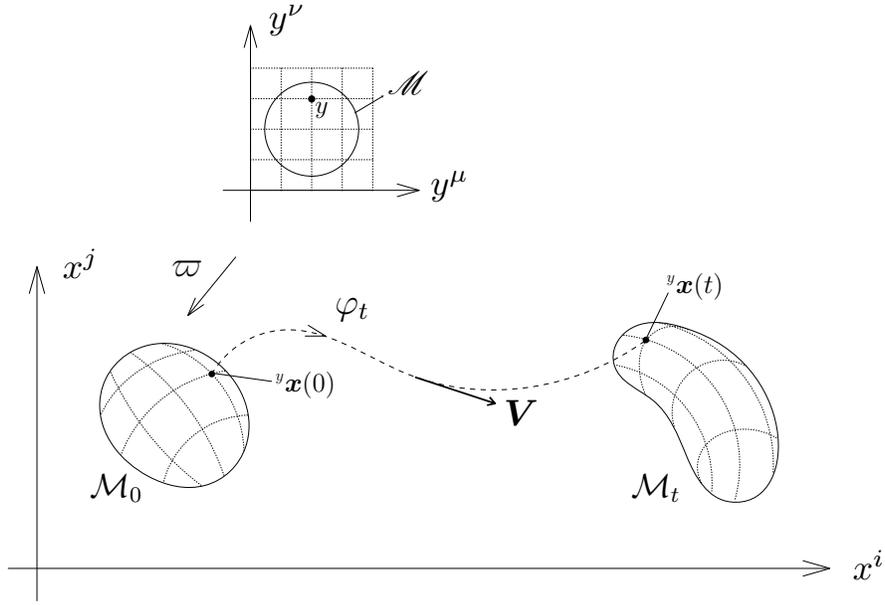}
\caption{An arbitrary open set occupied by fluid can be identified as a moving manifold $\mathcal{M}_t$ convected by the mean flow. We introduce the mean-Lagrangian coordinates $\{y^{{}_1},y^{{}_2},y^{{}_3}\}\in \mathscr{M}$ as labels identifying points on $\mathcal{M}_t$ convected by the mean velocity $\vc{V}$.}
\label{mean-Lagrangian picture}
\end{figure}

In the later discussions we will focus on the history effect in perturbation analysis, where the mean-Lagrangian picture of Ref. \cite{Ariki17} offers a suitable tool for generally-covariant description of the history effect. The mean-Lagrangian picture is a Lagrangian-type picture on the basis of the mean flow. Consider a point, say $\mlp$, whose trajectory may be written as $\mlpind\vc{x}(t)$ in a general coordinate system $\{x^{\scalebox{0.6}{1}},x^{\scalebox{0.6}{2}},x^{\scalebox{0.6}{3}}\}$. For a given mean velocity field $\vc{V}(\vc{x},t)$, the trajectory is determined by the following first-order differential equation:
\begin{equation}
\frac{d}{dt}\,\mlpind\vc{x}(t)=\vc{V}\left(\mlpind\vc{x}(t),t\right)
\end{equation}
with its initial position $\mlpind\vc{x}(0)$. As far as $\vc{v}$ is smooth, its ensemble average $\vc{V}$ is also a smooth vector field, and the trajectory $\mlpind\vc{x}(t)$ is uniquely determined for its initial position $\mlpind\vc{x}(0)$, which forms a 1-parameter group acting on spacetime:
\begin{equation}
\varphi_t: \left(\mlpind\vc{x}(0),0\right)
\mapsto \left( \mlpind\vc{x}(t),t\right),
\end{equation}
Then, assembly of the point $\mlpind\vc{x}(t_0)$, an open set $\mathcal{M}_0$, is recognized as a moving manifold $\mathcal{M}_t=\varphi_t \mathcal{M}_{0}$. Consider another open set $\mathscr{M}\in\mathbb{R}^3$. With the help of a diffeomorphism $\varpi: \mathscr{M}\to\mathcal{M}_0$, $\mathscr{M}$ acts as a label identifying these moving points;
\begin{equation}
\mathscr{M}\overset{\varpi}{\longrightarrow}\mathcal{M}_0\overset{\varphi_t}{\longrightarrow}\mathcal{M}_t
\end{equation}
We call $(y^{\scalebox{0.6}{1}},y^{\scalebox{0.6}{2}},y^{\scalebox{0.6}{3}})(\in\mathscr{M})$ the \emph{mean-Lagrangian coordinate system} (see Fig. \ref{mean-Lagrangian picture}). We treat the physics on the basis of the mean-Lagrangian coordinate space $\mathscr{M}$, where all the physical fields can be pulled back from the physical space. Then $\mathscr{M}$ is recognized as a metric space $\left(\,\mathscr{M}_t,g(t)\,\right)$ with a dynamical metric tensor $g_{\mu\nu}(\vc{y},t)$. For further details of the mean-Lagrangian formalism, the author refers the readers to Ref. \cite{Ariki17}. In the mean-Lagrangian coordinate system, we have the following set of governing equations:
\begin{equation}
\begin{split}
&\left( \frac{\partial}{\partial t}-\nu\Delta\right)v'^\mu(\vc{y},t)+\left(v'^\mu v'^\rho \right)_{;\rho}(\vc{y},t)+\mathrm{p}'^{;\mu}(\vc{y},t)\\
&=-\left(S^\mu_\rho+{\Theta^\mu}_\rho\right)(\vc{y},t)\ v'^\rho(\vc{y},t)+{R^{\mu\rho}}_{;\rho}(\vc{y},t),
\end{split}
\label{v' eq. 2}
\end{equation}
\begin{equation}
{v'^\rho}_{;\rho}(\vc{y},t)=0,
\label{incomp 2}
\end{equation}
where we employ the Greek indices for the mean-Lagrangian representation.

\subsection{Two-scale representation}\label{TWO-SCALE REPRESENTATION}
\begin{figure}
\centering
\includegraphics[width=8cm]{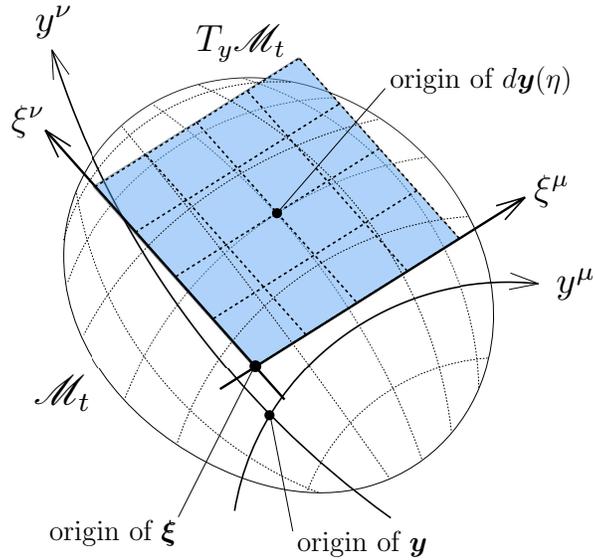}
\caption{$\vc{\xi}$-coordinates of a tangent space $T_y\mathscr{M}_t$ is shifted from the natural coordinates $dy^\mu(\eta)$ by $y^\mu$.
}
\label{xi frame}
\end{figure}
Here we apply the two-scale technique of Ref. \cite{Yoshizawa84} to Eqs. (\ref{v' eq. 2}) and (\ref{incomp 2}) in the mean-Lagrangian coordinate space $\mathscr{M}_t$. For this sake, we utilize the tangent bundle $T\mathscr{M}_t$. Let $\eta$ be a tangent vector on $\mlp(\in\mathscr{M}_t)$, i.e. $\eta \in T_y\mathscr{M}_t$. The set of coordinate variables $y^\mu \ (\mu=1,2,3)$ and its derivatives $\eta^\mu=dy^\mu(\eta)$ are often employed as natural coordinates of the tangent bundle $T \mathscr{M}_t$. In the present work, on the other hand, we employ $\xi^\mu(\mlp,\eta) \equiv dy^\mu(\eta)+y^\mu$ and $Y^\mu(\mlp,\eta) \equiv \delta y^\mu$ as another set of coordinate variables representing the point $(\mlp,\eta)$ on $T\mathscr{M}_t$, where $\delta$ is a bookkeeping parameter ($\delta=1$) introduced for later perturbation analysis. Then the coordinates $\xi^\mu$ of the tangent space $T_y\mathscr{M}_t$ are shifted from the natural coordinates $dy^\mu(\eta)$ by $y^\mu$ (see Fig. \ref{xi frame}). With this coordinate settings, we introduce two-scale representation of the dynamical field in a parallel manner to the original multiple-scale technique; for an arbitrary dynamical field $f(\vc{y},t)$ on $\mathscr{M}_t$, we introduce a corresponding function $f(\vc{\xi},t|\vc{Y})$ on the tangent bundle $T\mathscr{M}_t$. We call this correspondence \emph{two-scale representation} or simply TS:
\begin{equation*}
\mathrm{TS}: f(\vc{y},t) \longmapsto f(\boldsymbol{\xi},t|\vc{Y}).
\end{equation*}
We apply TS to all the dynamical field variables on the basis of the mean-Lagragian coordinates. Following two assumptions are setup:
\begin{enumerate}
\item The original field $f(\vc{y},t)$ on $\mathscr{M}_t$ is obtained from the TS variable $f(\boldsymbol{\xi},t|\vc{Y})$ on $T\mathscr{M}_t$ by substituting $\vc{\xi}=\vc{y}$ and $\vc{Y}=\delta\vc{y}$: $f(\vc{y},t|\delta\vc{y})=f(\vc{y},t)$.\\
\item An arbitrary $f(\boldsymbol{\xi},t|\vc{Y})$ is statistically homogeneous in $\boldsymbol{\xi}$.
\end{enumerate}
The first assumption is almost equivalent to what is assumed in multiple-scale expansion techniques. The second assumption is recursively validated in later discussion. TS should be consistently applied to both addition and multiplication of dynamical variables. This requires TS to conserve addition and multiplication:
\begin{equation*}
(f_1+f_1)(\vc{y},t)=f_1(\vc{y},t)+f_2(\vc{y},t)
\TS
(f_1+f_2)(\boldsymbol{\xi},t|\vc{Y})=f_1(\boldsymbol{\xi},t|\vc{Y})+f_2(\boldsymbol{\xi},t|\vc{Y}),
\end{equation*}
\begin{equation*}
(f_1 f_2)(\vc{y},t)=f_1(\vc{y},t)f_2(\vc{y},t)
\TS(f_1f_2)(\boldsymbol{\xi},t|\vc{Y})=f_1(\boldsymbol{\xi},t|\vc{Y})f_2(\boldsymbol{\xi},t|\vc{Y}). 
\end{equation*}
Under the second assumption, an arbitrary second-order moment associated with the same $\vc{Y}$ may be written as
\begin{equation}
\langle f_1(\boldsymbol{\xi},t|\vc{Y})f_2(\boldsymbol{\xi}',t'|\vc{Y})\rangle
=C(\boldsymbol{\xi}-\boldsymbol{\xi}',t,t'|\vc{Y}),
\end{equation}
where $C$ is a function dependent on the difference of tangent-space coordinates ($\boldsymbol{\xi}-\boldsymbol{\xi}'$), two times ($t$ and $t'$), and the mean-Lagrangian coordinate ($\vc{Y}$). As a special case, we have 
\begin{equation}
\langle f_1(\vc{y},t) f_2(\vc{y},t')\rangle
\TS
\langle f_1(\boldsymbol{\xi},t|\vc{Y})f_2(\boldsymbol{\xi},t'|\vc{Y})\rangle
=C(\boldsymbol{0},t,t'|\vc{Y}),
\end{equation}
which does not depend on $\boldsymbol{\xi}$.\\
 
\textit{ lemma 1. TS of one-point statistics is free from $\vc{\xi}$.}\\ 

\noindent Thus, for instance, TS of the Reynolds stress $\mathsfbi{R}$, mean strain rate $\mathsfbi{S}$, and mean absolute vorticity $\mathsfbi{\Theta}$ read 
\begin{equation*}
\mathsfbi{R}(\vc{y},t), \mathsfbi{S}(\vc{y},t), \mathsfbi{\Theta}(\vc{y},t)
\overset{{}^{\textrm{TS}}}{\longmapsto}
\mathsfbi{R}(t|\vc{Y}), \mathsfbi{S}(t|\vc{Y}), \mathsfbi{\Theta}(t|\vc{Y}).
\end{equation*}
For arbitrary tensor fields, we shall pay special attention to their duality; owing to the metric 2-form $ds^2=g_{\mu\nu}dy^\mu dy^\nu$, an arbitrary tensor fields on $T\mathscr{M}_t^{\otimes r}\otimes T^*\mathscr{M}_t^{\otimes s}$ have its dual on $T\mathscr{M}_t^{\otimes s}\otimes T^*\mathscr{M}_t^{\otimes r}$. For instance, we consider a contravariant vector $h^\mu(\vc{y},t)$. According to \emph{lemma 1}, TS of its moment is given by
\begin{equation}
\langle h^\mu(\vc{y},t)h^\nu(\vc{y},t)\rangle
=H^{\mu\nu}(\vc{y},t)\overset{{}^{\textrm{TS}}}{\longmapsto}
H^{\mu\nu}(t|\vc{Y})
\end{equation}
On the other hand, using covariant component $h_\mu(\vc{y},t)=g_{\mu\rho}(\vc{y},t)h^\rho(\vc{y},t)$, we obtain
\begin{equation}
\langle h_\mu(\vc{y},t)h_\nu(\vc{y},t)\rangle
=H_{\mu\nu}(\vc{y},t)\overset{{}^{\textrm{TS}}}{\longmapsto}
H_{\mu\nu}(t|\vc{Y})
\end{equation}
Using $h_\mu(\vc{y},t)\TS g_{\mu\rho}(\vc{\xi},t|\vc{Y})h^\rho(\vc{\xi},t|\vc{Y})$, we reach
\begin{equation}
H_{\mu\nu}(t|\vc{Y})
=g_{\mu\rho}(\vc{\xi},t|\vc{Y})g_{\nu\sigma}(\vc{\xi},t|\vc{Y})H^{\rho\sigma}(t|\vc{Y}).
\end{equation}
Comparing arguments of the both sides, we reach 
\begin{equation}
g_{\mu\nu}(\vc{y},t)\TS g_{\mu\nu}(t|\vc{Y})
\end{equation} 
(and equally $g^{\mu\nu}(\vc{y},t)\TS g^{\mu\nu}(t|\vc{Y})$).\\

\textit{Lemma 2. TS of metric is free from $\vc{\xi}$.}\\

\noindent This gives a flat metric to $T_y\mathscr{M}_t$, i.e. $d(\boldsymbol{\xi},\boldsymbol{\xi}')=g_{\mu\nu}(t|\vc{Y})(\xi^\mu-\xi'^\mu)(\xi^\nu-\xi'^\nu)$. Regarding an identity $\partial_t g_{\mu\nu}(\vc{y},t)=S_{\mu\nu}(\vc{y},t)$, \emph{lemma 2} gives the corresponding relation in TS: $\partial_t g_{\mu\nu}(t|\vc{Y})=S_{\mu\nu}(t|\vc{Y})$. According to \emph{lemma 2}, we can introduce Levi-Civita connection in TS representation. The Christoffel symbol of the first kind reads
\begin{equation}
\Gamma_{\rho.\mu\nu}\equiv
\frac{1}{2}\left(
\frac{\partial}{\partial y^\nu}g_{\rho\mu}
+\frac{\partial}{\partial y^\mu}g_{\rho\nu}
-\frac{\partial}{\partial y^\rho}g_{\mu\nu}\right)
\TS
\frac{1}{2}\delta\left(
\frac{\partial}{\partial Y^\nu}g_{\rho\mu}
+\frac{\partial}{\partial Y^\mu}g_{\rho\nu}
-\frac{\partial}{\partial Y^\rho}g_{\mu\nu}\right),
\end{equation}
which may be simply written as $\delta\Gamma_{\rho.\mu\nu}(t|\vc{Y})$. Obviously, its second kind becomes 
\begin{equation}
\Gamma^\rho_{\mu\nu}(\vc{y},t)=g^{\rho\sigma}(\vc{y},t)\Gamma_{\sigma.\mu\nu}(\vc{y},t)
\TS
g^{\rho\sigma}(t|\vc{Y})\delta\Gamma_{\rho.\mu\nu}(t|\vc{Y})=\delta\Gamma^\rho_{\mu\nu}(t|\vc{Y}).
\end{equation}
Using the connection field, derivative operation can be introduced for TS variables; TS of the simple spatial derivative follows from the first assumption:
\begin{equation}
\partial/\partial y^\mu \TS \partial/\partial \xi^\mu+\delta \partial/\partial Y^\mu,
\end{equation}
while the covariant derivative becomes
\begin{equation}
\nabla_\kappa {T^{\mu\nu\dots}}_{\rho\sigma\dots}(\vc{y},t)
\TS
\left( \frac{\partial}{\partial \xi^\kappa} + \delta\ \yna_\kappa \right)
{T^{\mu\nu\dots}}_{\rho\sigma\dots}(\boldsymbol{\xi},t|\vc{Y}),
\end{equation}
where the operator $\yna$, which may be called the covariant derivative by $\vc{Y}$ or simply the covariant derivative if not confusing, is given by
\begin{equation}
\begin{split}
\yna_\kappa {T^{\mu\nu\dots}}_{\rho\sigma\dots}(\boldsymbol{\xi},t|\vc{Y})
=&\frac{\partial}{\partial Y^\kappa}
{T^{\mu\nu\dots}}_{\rho\sigma\dots}(\boldsymbol{\xi},t|\vc{Y})\\
 &+\Gamma^\mu_{\alpha\kappa}(t|\vc{Y}){T^{\alpha\nu\dots}}_{\rho\sigma\dots}(\boldsymbol{\xi},t|\vc{Y})
 +\Gamma^\nu_{\alpha\kappa}(t|\vc{Y}){T^{\mu\alpha\dots}}_{\rho\sigma\dots}(\boldsymbol{\xi},t|\vc{Y})+\cdots\\
&-\Gamma^\beta_{\rho\kappa}(t|\vc{Y}) {T^{\mu\nu\dots}}_{\beta\sigma\dots}(\boldsymbol{\xi},t|\vc{Y})
-\Gamma^\beta_{\sigma\kappa}(t|\vc{Y}) {T^{\mu\nu\dots}}_{\rho\beta\dots}(\boldsymbol{\xi},t|\vc{Y})-\cdots\ . 
\end{split}
\label{yna}
\end{equation}
By definition, our connection is metrical and satisfies a known identity $\yna_\rho g_{\mu\nu}(t|\vc{Y})=0$.\\

Now \emph{Lemmas 1} and \emph{2} and all their consequences Eqs. (2.8)-(2.18) offer sufficient properties of our TS. By applying TS to Eqs. (\ref{v' eq. 2}) and (\ref{incomp 2}) we obtain a closed set of equations for TS variables $\vc{v}'(\boldsymbol{\xi},t|\vc{Y})$ and $\mathrm{p}'(\boldsymbol{\xi},t|\vc{Y})$ on tangent bundle $T\mathscr{M}_t$:
\begin{equation}
\begin{split}
&\bigg\{\frac{\partial}{\partial t}-\nu g^{\rho\sigma}(t|\vc{Y})
\frac{\partial}{\partial \xi^\rho}\frac{\partial}{\partial \xi^\sigma}
-2\delta\nu\frac{\partial}{\partial \xi^\rho}\yna^\rho-\delta^2\nu\yna^2\bigg\}
v'^\mu(\boldsymbol{\xi},t|\vc{Y})\\
&+\left(\frac{\partial}{\partial\xi^\rho}+\delta\,\yna_\rho\right)
(v'^\mu v'^\rho)(\boldsymbol{\xi},t|\vc{Y})+g^{\mu\nu}(t|\vc{Y})\left(\frac{\partial}{\partial\xi^\nu}+\delta\,\yna_\nu\right)\mathrm{p}'(\boldsymbol{\xi},t|\vc{Y})\\
=&-\left(S^\mu_\rho+{\Theta^\mu}_\rho\right)(t|\vc{Y})v'^\rho(\boldsymbol{\xi},t|\vc{Y})
+\delta\,\yna_\rho R^{\mu\rho}(t|\vc{Y}),
\end{split}
\label{MLTS eq}
\end{equation}
\begin{equation}
\left(\frac{\partial}{\partial \xi^\nu}+\delta\,\yna_\nu \right)v'^\nu(\boldsymbol{\xi},t|\vc{Y})=0.
\label{MLTS incompressibility}
\end{equation}
We obtain the solutions $\vc{v}'(\vc{y},t)$ and $\mathrm{p}'(\vc{y},t)$ of the original Eqs. (\ref{v' eq. 2})-(\ref{incomp 2}) by substituting $\vc{\xi}=\vc{y}$ and $\vc{Y}=\delta \vc{y}$ into $\vc{v}'(\boldsymbol{\xi},t|\vc{Y})$ and $\mathrm{p}'(\boldsymbol{\xi},t|\vc{Y})$. Here we should notice that Eqs. (\ref{MLTS eq}) and (\ref{MLTS incompressibility}) allow the solutions to be homogeneous with respect to $\boldsymbol{\xi}$, since their forms are identical under the translation;
\begin{equation}
\boldsymbol{\xi}\to \boldsymbol{\xi'}=\boldsymbol{\xi}+\mathbf{a},
\end{equation}
where $\mathbf{a}$ is a constant vector independent of $\boldsymbol{\xi}$ and $t$. In the present formulation, all the fluctuating quantities are constructed from $\vc{v}'$ and $\mathrm{p}'$ both of which are allowed to be statistically homogeneous in $\boldsymbol{\xi}$, which may validate the second assumption.\\

More intuitive picture of TS representation may be understood from the scale-separation concept  likewise in TSDIA \cite{Yoshizawa84}; under weak inhomogeneity, TS variables may express fine-scale variations of fluctuations by $\vc{\xi}$ and large-scale behaviors of the mean fields by $\vc{Y}$. 

It is also worthwhile to compare our TS representation with conventional methodologies of Refs. \citep{Leslie73,Besnard96}. Providing a two-point correlation $\mathscr{C}(\vc{x}_{\scalebox{0.6}{$(1)$}},\vc{x}_{\scalebox{0.6}{$(2)$}})$ defined at two spatial points $\vc{x}_{\scalebox{0.6}{$(1)$}}$ and $\vc{x}_{\scalebox{0.6}{$(2)$}}$ in a Cartesian coordinate setting, one can introduce two-type of coordinate variables, i.e. the difference coordinate $\vc{r}\equiv(\vc{x}_{\scalebox{0.6}{$(1)$}}-\vc{x}_{\scalebox{0.6}{$(2)$}})$ and the centroid coordinate $\vc{x}\equiv(\vc{x}_{\scalebox{0.6}{$(1)$}}+\vc{x}_{\scalebox{0.6}{$(2)$}})/2$, and rewrite the correlation as $\mathscr{C}(\vc{x}+\vc{r}/2,\vc{x}-\vc{r}/2)$. Regarding $\vc{r}$ as the coordinate variable describing fine-scale variations, one may approximate $\mathscr{C}(\vc{x}+\vc{r}/2,\vc{x}-\vc{r}/2)$ to be homogeneous in $\vc{r}$, while the inhomogeneity accounted by another coordinate $\vc{x}$ representing large-scale variations of the mean field. Then $\vc{r}$ and $\vc{x}$ correspond respectively to $\vc{\xi}$ and $\vc{Y}$ of the present TS representation, although their coordinate-decomposition techniques can hardly be applied to general curvilinear coordinates just like we did in this section.

\subsection{Fourier transformation}\label{FOURIER TRANSFORM}
On the basis of the metric tensor, we introduce the cotangent space $T^*_y\mathscr{M}$ as the wavenumber space. Then, using the Fourier transform, an arbitrary TS-function $f(\boldsymbol{\xi},t|\vc{Y})$ on the tangent bundle $T\mathscr{M}_t$ can be converted to a function $f(\vc{k},t|\vc{Y})$ on the cotangent bundle $T^*\mathscr{M}_t$. We define the Fourier transformation $\mathcal{F}:T\mathscr{M}_t\rightarrow T^*\mathscr{M}_t$ and its inverse $\mathcal{F}^{-1}:T^*\mathscr{M}_t\rightarrow T\mathscr{M}_t$ by 
\begin{subequations}
\begin{align}
&\mathcal{F}\times \equiv (2\pi)^{-3} \int dvol_\xi \exp(-ik_\rho\xi^\rho)\times,\\
&\mathcal{F}^{-1}\times \equiv \int dvol_k \exp(ik_\rho\xi^\rho)\times,
\end{align}
\label{Fourier}
\end{subequations}
where $dvol_\xi=\sqrt{\mathcal{G}}d\xi^{\scalebox{0.6}{1}}\wedge d\xi^{\scalebox{0.6}{2}}\wedge d\xi^{\scalebox{0.6}{3}}$ and $dvol_k=\sqrt{\mathcal{G}^{-1}}dk_{\scalebox{0.6}{1}}\wedge dk_{\scalebox{0.6}{2}}\wedge dk_{\scalebox{0.6}{3}}$ ($\mathcal{G}\equiv\det[g_{\rho\sigma}]$) are the invariant-volume forms in $T_y\mathscr{M}_t$ ($\vc{\xi}$ space) and $T^*_y\mathscr{M}_t$ ($\vc{k}$ space) respectively. Under this definition, covariant delta function in wavenumber space is given by
\begin{equation}
\delta^3_c(\vc{k}|\vc{Y})=(2\pi)^{-3} \int dvol_\xi \exp(-ik_\rho\xi^\rho)=\sqrt{\mathcal{G}}\delta^3(\vc{k}).
\label{covariant delta}
\end{equation}
Nonlinear term is transformed into the convolution:
\begin{equation*}
\cnv{k}{p}{q}f(\vc{p},t|\vc{Y})g(\vc{q},t|\vc{Y})\equiv\int dvol_p \int dvol_q \delta^3_c(\vc{k}-\vc{p}-\vc{q}|\vc{Y})f(\vc{p},t|\vc{Y})g(\vc{q},t|\vc{Y}).
\end{equation*}

It is important to see the commutation relations between the Fourier transformation and derivative operations. For this sake, we investigate $\mathcal{G}$ in $\mathcal{F}$, the only factor dependent on $\vc{Y}$ and $t$. Taking the covariant derivative of $\mathcal{G}$, we obtain
\begin{equation}
\yna_\rho \mathcal{G}(t|\vc{Y})=\mathcal{G}(t|\vc{Y})g^{\alpha\beta}(t|\vc{Y})\,\yna_\rho \,g_{\alpha\beta}(t|\vc{Y})=0,
\end{equation}
where we used a known formula $\partial\ \mathrm{det}[M_{\mu\nu}]/\partial M_{\alpha\beta} = \mathrm{det}[M_{\mu\nu}]\,(M^{-1})^{\alpha\beta}$ for an arbitrary two-rank matrix $\boldsymbol{\mathsf{M}}$. In the same manner, taking the time derivative of $\mathcal{G}$ yields
\begin{equation}
\begin{split}
\frac{\partial}{\partial t}\mathcal{G}(t|\vc{Y})&=\mathcal{G}(t|\vc{Y})g^{\alpha\beta}(t|\vc{Y})\frac{\partial}{\partial t}g_{\alpha\beta}(t|\vc{Y})\\
&=\mathcal{G}(t|\vc{Y})g^{\alpha\beta}(t|\vc{Y})S_{\alpha\beta}(t|\vc{Y})=0,
\end{split}
\end{equation}
for incompressibility. Thus the Fourier transformation $\mathcal{F}$ (and also $\mathcal{F}^{-1}$) commutes with both spatial and time derivatives \cite{TSDIA commutation}:
\begin{subequations}
\begin{align*}
\yna\ \mathcal{F}-\mathcal{F}\ \yna&=\yna\ \mathcal{F}^{-1}-\mathcal{F}^{-1}\ \yna=0,\\
\partial_t\ \mathcal{F}-\mathcal{F}\ \partial_t&=\partial_t\ \mathcal{F}^{-1}-\mathcal{F}^{-1}\ \partial_t=0.
\end{align*}
\end{subequations}
Finally, by applying the Fourier transformation to both Eqs. (\ref{MLTS eq}) and (\ref{MLTS incompressibility}), we obtain the closed set of equations for $\vc{v}'(\vc{k},t|\vc{Y})$ and $\mathrm{p}'(\vc{k},t|\vc{Y})$:
\begin{equation}
\begin{split}
&\bigg\{\frac{\partial}{\partial t}+\nu k^2-2i\delta\nu k_\rho\,\yna^\rho -\delta^2\,\yna^2 \bigg\}
v'^\mu(\vc{k},t|\vc{Y})\\
&+\left( ik_\nu+\delta\,\yna_\nu\right)
\cnv{k}{p}{q}v'^\mu(\vc{p},t|\vc{Y})v'^\nu(\vc{q},t|\vc{Y})+\left( ik^\mu+\delta\,\yna^\mu\right)\mathrm{p}'(\vc{k},t|\vc{Y})\\
&+\left( S^\mu_\nu+{\Theta^\mu}_\nu \right)(t|\vc{Y})v'^\nu(\vc{k},t|\vc{Y})
-\delta\,\yna R^{\mu\nu}(t|\vc{Y})\delta^3_c(\vc{k}|\vc{Y})\\
&=0,
\end{split}
\label{MLTS eq 2}
\end{equation}
\begin{equation}
\left( ik_\rho +\delta\,\yna_\rho \right)v'^\rho(\vc{k},t|\vc{Y})=0,
\label{MLTS incomp 2}
\end{equation}
where $k^2=g^{\rho\sigma}(t|\vc{Y})k_\rho k_\sigma$.

\subsection{Solenoidal projection}\label{PRESSURE ELIMINATION}
In Eqs. (\ref{MLTS eq 2}) and (\ref{MLTS incomp 2}) we have two dynamical variables: $\vc{v}'(\vc{k},t|\vc{Y})$ and $\mathrm{p}'(\vc{k},t|\vc{Y})$. In case of homogeneous turbulence, one can eliminate the pressure simply by multiplying the Navier-Stokes equation by the solenoidal operator, which greatly simplifies the later analysis. Following a similar step, we can eliminate the pressure $\mathrm{p}'(\vc{k},t|\vc{Y})$ from Eq. (\ref{MLTS eq 2}) using a modified solenoidal operator:
\begin{equation}
\begin{split}
\hat{P}^\mu_\nu
=P^\mu_\nu-\delta P^\mu_\rho\,\yna^\rho\,
{}^{\scalebox{0.5}{p}}\!\hat{L}^{-1}\frac{k_\nu}{ik^2},
\end{split}
\label{modified projection}
\end{equation}
where $P^\mu_\nu=\delta^\mu_\nu-k^\mu k_\nu/k^2$, ${}^{\scalebox{0.5}{p}}\!\hat{L}=1-\delta ik_\rho \,\yna^\rho/k^2$, and $k=\sqrt{g^{\alpha\beta}k_\alpha k_\beta}$. Now the projection operator $\hat{P}^\mu_\nu$ cancels out the first-order derivative operation $ik^\mu+\delta\yna^\nu$ i.e. $\hat{P}^\mu_\nu\left( ik^\nu+\delta\,\yna^\nu\right)=0$, just like a wellknown operation $P^\mu_\nu k^\nu=0$. Then the pressure related term vanishes from Eq. (\ref{MLTS eq 2}) by multiplying its both sides by $\hat{\mathsfbi{P}}$ (see appendix \ref{Derive expanded eq} for the derivation of $\hat{\mathsfbi{P}}$):
\begin{equation}
\hat{P}^\mu_\nu\ \mathscr{U}^\nu\left[\vc{v}'\right]=0,
\label{MLTS eq 3}
\end{equation}
where $\mathscr{U}^\nu\left[\vc{v}'\right]$ is a functional of $\vc{v}'$ given by the left side of Eq. (\ref{MLTS eq 2}) without the pressure term (also see Eq. (\ref{U[v]}) in appendix \ref{Derive expanded eq}).\\

In the later analysis, we will take the similar procedure to that of the incompressible-turbulence theory. Thus it is useful to introduce the solenoidal (incompressible) part of the velocity fluctuation. Following Ref. \cite{Hamba87}, we introduce
\begin{equation}
\sol v^\mu(\vc{k},t|\vc{Y})=P^\mu_\nu v'^\nu(\vc{k},t|\vc{Y}),
\end{equation}
where $P^\mu_\nu=\delta^\mu_\nu-k^\mu k_\nu /k^2$ is a projection operator for extracting the solenoidal part. Using Eq. (\ref{MLTS incomp 2}), we obtain
\begin{equation}
\begin{split}
\sol v^\mu(\vc{k},t|\vc{Y})
&=\left(\delta^\mu_\nu-\frac{k^\mu k_\nu}{k^2}\right) v'^\nu(\vc{k},t|\vc{Y})\\
&=v'^\mu(\vc{k},t|\vc{Y})-\delta\frac{ik^\mu\ \yna_\nu}{k^2}v'^\nu(\vc{k},t|\vc{Y})\\
&=\sol\hat{L}^\mu_\nu v'^\nu(\vc{k},t|\vc{Y}),
\end{split}
\end{equation}
where $\sol\hat{L}^\mu_\nu=\delta^\mu_\nu-\delta ik^\mu\,\yna_\nu /k^2$. Using $\vc{v}'=\sol\hat{\mathsfbi{L}}{}^{\scalebox{0.7}{-1}}\ \sol\vc{v}$, Eq. (\ref{MLTS eq 3}) may be rewritten as those of $\sol\vc{v}$:
\begin{equation}
\hat{P}^\mu_\nu\ \mathscr{U}^\nu\left[ \sol\hat{\mathsfbi{L}}^{-1}\ \sol\vc{v}\right]=0,
\label{master eq.}
\end{equation}
Note that Eq. (\ref{MLTS incomp 2}) is always satisfied for $\vc{v}'$ given by $\sol\hat{\mathsfbi{L}}{}^{\scalebox{0.7}{-1}}\,\vc\sol\vc{v}$, so our governing equations (\ref{MLTS eq 2}) and (\ref{MLTS incomp 2}) are finally reduced to a single Eq. (\ref{master eq.}). For the actual calculation, we shall rely on perturbation analysis so we expand Eq. (\ref{master eq.}) in terms of $\delta$ (see appendix \ref{Derive expanded eq}):
\begin{equation}
\begin{split}
\left(\frac{\partial}{\partial t}+\nu k^2 \right)\,\sol v^\mu(\vc{k},t|\vc{Y})
=&\frac{1}{i}M^\mu_{\rho\sigma}
\cnv{k}{p}{q}
\,\sol v^\mu(\vc{p},t|\vc{Y})
\,\sol v^\mu(\vc{q},t|\vc{Y})\\
&-P^\mu_\nu
(S^\nu_\rho+{\Theta^\nu}_\rho)(t|\vc{Y})
\,\sol v^\rho(\vc{q},t|\vc{Y})\\
&+O(\delta)
\label{expanded eq}
\end{split}
\end{equation}
where $M^\mu_{\rho\sigma}=\frac{1}{2}(P^\mu_\rho k_\sigma+P^\mu_\sigma k_\rho)$.

\subsection{Orthonormal frame field}\label{ORTHONORMAL}
In the forthcoming Sec. \ref{PERTURBATION}-\ref{RENORMALIZATION}, we will apply perturbation analysis to Eq. (\ref{expanded eq}), where homogeneous-isotropic state is taken as the basic solution (see Sec. \ref{BASIC}). In this regard, orthonormal frame may provide suitable representation in treating isotropic state. An orthonormal frame field (vielbein) $e^\mu_I(t|\vc{Y})$ and coframe field (conbein) $\bar{e}^I_\mu(t|\vc{Y})$ ($I=1,2,3$) are defined by $e_I^\mu e_J^\nu g_{\mu\nu}=\delta_{IJ}$ and $\bar{e}^I_\mu \bar{e}^J_\nu g^{\mu\nu}=\delta^{IJ}$, giving the orthonormal metric by a linear transformation. In this frame representation, an arbitrary covariant (contravatiant) vector $k_\mu$ ($\xi^\mu$) is represented as $\check{k}_I=e^\mu_I k_\mu$ ($\check{\xi}^I=\bar{e}_\mu^I \xi^\mu$). Dynamical equations of frame and coframe fields follow from time derivatives of $e_I^\mu e_J^\nu g_{\mu\nu}=\delta_{IJ}$ and $\bar{e}^I_\mu \bar{e}^J_\nu g^{\mu\nu}=\delta^{IJ}$:
\begin{subequations}
\begin{align*}
&\left(\frac{\partial}{\partial t}e_I^\mu\right)e_J^\nu g_{\mu\nu}
+\bar{e}^I_\mu\left(\frac{\partial}{\partial t}e_J^\nu\right)g_{\mu\nu}
+e_I^\mu e_J^\nu\,S_{\mu\nu}=\frac{\partial}{\partial t}\delta^{IJ}=0,\\
&\left(\frac{\partial}{\partial t}\bar{e}^I_\mu\right)\bar{e}^J_\nu g^{\mu\nu}
+\bar{e}^I_\mu\left(\frac{\partial}{\partial t}\bar{e}^J_\nu\right)g^{\mu\nu}
-\bar{e}^I_\mu \bar{e}^J_\nu\,S^{\mu\nu}=\frac{\partial}{\partial t}\delta^{IJ}=0,
\end{align*}
\end{subequations}
which are symmetric-tensor equations for six components. An arbitrary antisymmetric tensor $\mathsfbi{A}$ can complement the above equations to be nine-component equations:
\begin{subequations}
\begin{align*}
&\left(\frac{\partial}{\partial t}e_I^\mu\right)e_J^\nu g_{\mu\nu}
=-\frac{1}{2}e_I^\mu e_J^\nu\,(S_{\mu\nu}+A_{\mu\nu}),\\
&\left(\frac{\partial}{\partial t}\bar{e}^I_\mu\right)\bar{e}^J_\nu g^{\mu\nu}
=\frac{1}{2}\bar{e}^I_\mu \bar{e}^J_\nu\,(S^{\mu\nu}+A^{\mu\nu}),
\end{align*}
\end{subequations}
which read, with arguments $t$ and $\vc{Y}$ explicitly given,
\begin{subequations}
\begin{align}
\partial_t e_J^\nu(t|\vc{Y})
&=-\frac{1}{2}(S^\nu_\mu+A_\mu{}^\nu)(t|\vc{Y})\,e_J^\mu(t|\vc{Y}),\\
\partial_t \bar{e}^I_\mu(t|\vc{Y})
&=\frac{1}{2}(S^\nu_\mu+A_\mu{}^\nu)(t|\vc{Y})\,\bar{e}^I_\nu(t|\vc{Y}).
\end{align}
\label{frame field eq}
\end{subequations}
The arbitrariness in $\mathsfbi{A}$ arises from $SO(3)$ symmetry of the orthonormal frame, which will be reconsidered in Sec. \ref{PERTURBATION}. Introducing the time-advancement coefficients $\Lambda^\nu_\mu(t,t'|\vc{Y})\equiv e^\nu_I(t|\vc{Y}) \bar{e}^I_\mu(t'|\vc{Y})$ and $\bar{\Lambda}^\nu_\mu(t,t'|\vc{Y})\equiv \bar{e}^J_\mu(t|\vc{Y}) e^\nu_J(t'|\vc{Y})$ ($\bar{\boldsymbol{\Lambda}}=\boldsymbol{\Lambda}^{-1}$), we express $\bar{e}^I_\mu(t|\vc{Y})=\bar{\Lambda}^\rho_\mu(t,t'|\vc{Y})\bar{e}^I_\rho(t'|\vc{Y})$ and $e^\nu_J(t|\vc{Y})=\Lambda^\nu_\rho(t,t'|\vc{Y})e^\rho_J(t'|\vc{Y})$, where
\begin{subequations}
\begin{align}
\partial_t \Lambda^\nu_\mu(t,t'|\vc{Y})
&=-\frac{1}{2}(S^\nu_\rho+A_\rho{}^\nu)(t|\vc{Y})\,\Lambda^\rho_\mu(t,t'|\vc{Y}),\\
\partial_t \bar{\Lambda}^\nu_\mu(t,t'|\vc{Y})
&=\frac{1}{2}(S^\rho_\mu+A_\mu{}^\rho)(t|\vc{Y})\,\bar{\Lambda}^\nu_\rho(t,t'|\vc{Y}),
\end{align}
\label{Lambda}
\end{subequations}
with $\boldsymbol{\Lambda}(t',t'|\vc{Y})=\bar{\boldsymbol{\Lambda}}(t',t'|\vc{Y})=\boldsymbol{1}$. In the orthonormal frame, we define a differential operator for the linearized Navier-Stokes equation:
\begin{equation}
\hat{L}^I_J=P^I_J\left(\frac{\partial}{\partial t}+\nu \check{k}^2\right),
\end{equation} 
where $\check{k}^2=\delta^{IJ}\check{k}_I\check{k}_J$. In the mean-Lagrangian representation, this becomes
\begin{equation}
\hat{L}^\mu_\nu
=P^\mu_\nu\left(\frac{\partial}{\partial t}+\nu k^2\right)
+\frac{1}{2}P^\mu_\rho(S^\rho_\nu+A_\nu{}^\rho)
+\frac{1}{2}P^\mu_\nu(S^\rho_\sigma+A_\sigma{}^\rho) k_\rho\frac{\partial}{\partial k_\sigma}.
\label{Lhat}
\end{equation}
Then we transform the dynamical equation (\ref{expanded eq}) into  
\begin{equation}
\begin{split}
\hat{L}^\mu_\nu\, \sol v^\nu(\vc{k},t|\vc{Y})
=&\frac{1}{i}M^\mu_{\rho\sigma}\cnv{k}{p}{q}
\,\sol v^\rho(\vc{p},t|\vc{Y})
\,\sol v^\sigma(\vc{q},t|\vc{Y})\\
&+\hat{F}^\mu_\nu(\vc{k},t|\vc{Y})
\,\sol v^\nu(\vc{k},t|\vc{Y})
+O(\delta),
\end{split}
\label{pre perturbative eq}
\end{equation}
where $\hat{\boldsymbol{\mathsf{F}}}$ is another linear operator: 
\begin{equation}
\begin{split}
\hat{F}^\mu_\nu=P^\mu_\rho\left( -\frac{1}{2}S^\rho_\nu-\Theta_\nu{}^\rho +\frac{1}{2}{A_\nu}^\rho\right)
+\frac{1}{2}P^\mu_\nu(S^\rho_\sigma+{A_\sigma}^\rho)k_\rho\frac{\partial}{\partial k_\sigma}.
\end{split}
\end{equation}
Now Eq. (\ref{pre perturbative eq}) is interpreted as a perturbed Navier-Stokes equation, regarding $\hat{\mathsfbi{F}}$- and $\delta$-related terms on the right side as perturbation. In the forth-comming sections, we will apply perturbation analyses on the basis of Eq. (\ref{pre perturbative eq}). The arbitrariness of the frame field, and equally of $\mathsfbi{A}$, behaves like a gauge symmetry, which do not alter the total dynamics governed by Eq. (\ref{expanded eq}), and equally by Eq. (\ref{vs eq}) in appendix \ref{Derive expanded eq}. It becomes, however, certainly important in perturbation analysis (see the last paragraph of Sec. \ref{PERTURBATION}). 

Finally, we mention the volume form in the orthonormal representation. A trivial identity $\sqrt{|g_{\mu\nu}(t)|}=\sqrt{|\delta_{IJ}e_\mu^I(t) e_\nu^J(t)|}=|\mathsfbi{e}|$ yields
\begin{equation}
dvol_k=\sqrt{\mathcal{G}}dk_{\scalebox{0.5}{1}}\wedge dk_{\scalebox{0.5}{2}}\wedge dk_{\scalebox{0.5}{3}}
=\sqrt{\mathcal{G}}|\mathsfbi{e}|^{-1} d\check{k}_{\scalebox{0.5}{1}}\wedge d\check{k}_{\scalebox{0.5}{2}}\wedge d\check{k}_{\scalebox{0.5}{3}}
=d\check{k}_{\scalebox{0.5}{1}}\wedge d\check{k}_{\scalebox{0.5}{2}}\wedge d\check{k}_{\scalebox{0.5}{3}}\equiv d^3\check{k}.
\end{equation}
Orthonormal frame offers a convenient platform to spherical integration in the Fourier space, which will be discussed in later Sec. \ref{space-time non-local expression}.

\subsection{Perturbative expansion}\label{PERTURBATION}
Our basic strategy is based on perturbative analysis. In this context we have to specify the perturbative terms, so that we introduce bookkeeping parameters $\lambda$ and $\chi$ ($\lambda=\chi=1$) for the nonlinear self-interaction and $\boldsymbol{\mathsf{S}}$, $\boldsymbol{\mathsf{\Theta}}$-related terms respectively. Thus we rewrite (\ref{pre perturbative eq}) as
\begin{equation}
\begin{split}
\hat{L}^\mu_\nu\,\sol v^\nu(\vc{k},t|\vc{Y})
=&\lambda\frac{1}{i}M^\mu_{\rho\sigma}\cnv{k}{p}{q}
\,\sol v^\rho(\vc{p},t|\vc{Y})
\,\sol v^\sigma(\vc{q},t|\vc{Y})\\
&+\chi\hat{F}^\mu_\nu(\vc{k},t|\vc{Y})\,\sol v^\nu(\vc{k},t|\vc{Y})\\
&+O(\delta).
\end{split}
\label{perturbative eq}
\end{equation}
We assume these three parameters $\lambda$, $\chi$ and $\delta$ as perturbative parameters representing the magnitude of nonlinearity, anisotropy and inhomogeneity respectively. After perturbation analyses we rewrite $\lambda$, $\chi$, $\delta$ $\rightarrow 1$. We regard the following solenoidal field $\tilde{\vc{v}}$ as the unperturbed field;
\begin{equation}
\hat{L}^\mu_\nu \tilde{v}^\nu(\vc{k},t|\vc{Y})=0.
\label{vtilde}
\end{equation}
In addition, we introduce the propagator of $\tilde{\vc{v}}$ as
\begin{equation}
\hat{L}^\mu_\rho \tilde{G}^\rho_\nu(\vc{k},t;\vc{k}',t'|\vc{Y})=0\  (t\geq t'),
\label{Gtilde}
\end{equation}
with an initial condition $\tilde{G}^\mu_\nu(\vc{k},t';\vc{k}',t'|\vc{Y})=P^\mu_\nu\delta^3_c(\vc{k}-\vc{k}'|\vc{Y})$. $\tilde{\vc{v}}$ and $\tilde{\boldsymbol{\mathsf{G}}}$ are to be referred to as the bare field and bare propagator. Using $\tilde{\boldsymbol{\mathsf{G}}}$, we can integrate Eq. (\ref{perturbative eq}) as
\begin{equation}
\begin{split}
\sol v^\mu(\vc{k},t|\vc{Y})
=&\tilde{v}^\mu(\vc{k},t|\vc{Y})\\
&+\lambda\int^t_{-\infty} dt'\int dvol_{k'} \tilde{G}^\mu_\nu(\vc{k},t;\vc{k}',t'|\vc{Y}) \frac{1}{i}M^\nu_{\rho\sigma}\cnv{k'}{p}{q}
\,\sol v^\rho(\vc{p},t|\vc{y})
\,\sol v^\sigma(\vc{q},t|\vc{Y})\\
&+\chi\int^t_{-\infty} dt'\int dvol_{k'}\tilde{G}^\mu_\nu(\vc{k},t;\vc{k}',t'|\vc{Y})
\hat{F}^\nu_\rho\,\sol v^\rho(\vc{k}',t'|\vc{Y})\\
&+O(\delta).
\end{split}
\label{expanded vs 0}
\end{equation}
By substituting the above expansion into $\sol\vc{v}$ on its right side iteratively, we obtain the solution of Eq. (\ref{perturbative eq}) as a series expanded by $\tilde{\vc{v}}$. Using the reverse expansion $\sol\hat{\mathsfbi{L}}{}^{\scalebox{0.7}{-1}}=\mathsfbi{1}+\delta i\vc{k}\otimes \vc{\yna}/k^2+O(\delta^2)$, we obtain the perturbative expansion of the total velocity fluctuation:
\begin{equation}
\begin{split}
v^\mu(\vc{k},t|\vc{Y})&=(\sol\hat{L}^{-1}){}^\mu_\nu\ \sol v^\nu(\vc{k},t|\vc{Y})\\
&=\sol v^\mu(\vc{k},t|\vc{Y})+\delta\frac{ik^\mu}{k^2}\yna_\nu \ \sol v^\nu(\vc{k},t|\vc{Y})
+O(\delta^2).
\end{split}
\end{equation}
Now we should provide $\delta\ \vc{\yna}\otimes\sol\vc{v}$ some proper physical interpretation. The dynamical equation of $\delta\ \vc{\yna}\otimes\sol\vc{v}$ reads
\begin{equation}
\begin{split}
\hat{L}^\mu_\nu\, \delta \yna_\alpha\, \sol v^\nu(\vc{k},t|\vc{Y})
=&2\lambda\delta\frac{1}{i}M^\mu_{\rho\sigma}\cnv{k}{p}{q}
\,\sol v^\rho(\vc{p},t|\vc{Y})
\,\yna_\alpha\ \sol v^\sigma(\vc{q},t|\vc{Y})\\
&+\delta\chi\hat{F}^\mu_{\nu}\ \yna_\alpha\ \sol v^\nu(\vc{k},t|\vc{Y})\\
&+\delta\chi P^\mu_\rho\left(S^\rho_{\sigma;\alpha}+\Theta^\rho{}_{\sigma;\alpha}\right)(t|\vc{Y})\,\sol v^\sigma(\vc{k},t|\vc{Y}),
\end{split}
\end{equation}
which can be solved iteratively as
\begin{equation}
\begin{split}
\delta \yna_\alpha\,\sol &v^\nu(\vc{k},t|\vc{Y})\\
=&\delta\ \yna_\alpha\tilde{v}^\nu(\vc{k},t|\vc{Y})\\
&+\delta\chi \int^t_{-\infty} dt'\int dvol_{k'} \tilde{G}^\mu_\rho(\vc{k},t;\vc{k}',t'|\vc{Y})
\hat{F}^\rho_{\sigma}\ \yna_\alpha\,\tilde{v}^\sigma(\vc{k}',t'|\vc{Y})\\
&+\delta\chi\int^t_{-\infty} dt'\, \int dvol_{k'}\tilde{G}^\mu_\rho(\vc{k},t;\vc{k}',t'|\vc{Y})\left(S^\rho_{\sigma;\alpha}+\Theta^\rho{}_{\sigma;\alpha}\right)(t'|\vc{Y})\,\tilde{v}^\sigma(\vc{k}',t'|\vc{Y})\\
&+O(\lambda,\delta^2).
\end{split}
\label{delta perturbation 0}
\end{equation}
Here the first term ($\delta\ \vc{\yna}\otimes\tilde{\vc{v}}$) may reflect non-uniform distribution of unperturbed field. Now Eq. (\ref{delta perturbation 0}) allows us to understand $\delta$ effect from two aspects; one is caused by nonuniform distribution of unperturbed field in past, and the other by the inhomogeneity of $\mathsfbi{S}$ and $\mathsfbi{\Theta}$. Reminded that turbulence is primarily enhanced by turbulence production process related to $\mathsfbi{S}$ and $\mathsfbi{\Theta}$, here we focus only on $\delta$ effect caused by inhomogeneity of $\mathsfbi{S}$ and $\mathsfbi{\Theta}$ (for $\delta$-effect cased by $\delta\ \vc{\yna}\otimes\tilde{\vc{v}}$, see appendix \ref{DELTA}). Considering $\mathsfbi{S}$ and $\mathsfbi{\Theta}$ causes departure from isotropic fluctuation in Eq. (\ref{expanded vs 0}), their derivatives may well causes further deviation from the homogeneous isotropic state: 
\begin{equation}
\begin{split}
\delta\ \yna_\alpha\, \sol &v^\nu(\vc{k},t|\vc{Y})\\
=&\delta\chi\int^t_{-\infty} dt'\ \int dvol_{k'} \tilde{G}^\mu_\rho(\vc{k},t;\vc{k}',t'|\vc{Y})
\left(S^\rho_{\sigma;\alpha}+\Theta^\rho{}_{\sigma;\alpha}\right)(t'|\vc{Y})\,\tilde{v}^\sigma(\vc{k}',t'|\vc{Y})\\
&+O(\lambda,\delta^2).
\end{split}
\end{equation}
Now $\delta\ \vc{\yna}\otimes\vc{v}$ can be expanded in terms of $\tilde{v}^\sigma(\vc{k},t'|\vc{Y})$, which enables us to expand $\vc{v}'$ in terms of $\tilde{\vc{v}}$ and $\tilde{\mathsfbi{G}}$:
\begin{equation}
\begin{split}
v'^\mu&(\vc{k},t|\vc{Y})\\
=&\tilde{v}^\mu(\vc{k},t|\vc{Y})\\
&+\lambda\int^t_{-\infty} dt'\int dvol_{k'} \tilde{G}^\mu_\nu(\vc{k},t;\vc{k}',t'|\vc{Y})\frac{1}{i}M^\nu_{\rho\sigma}\cnv{k'}{p}{q}
\,\tilde{v}^\rho(\vc{p},t|\vc{Y})
\,\tilde{v}^\sigma(\vc{q},t|\vc{Y})\\
&+\chi\int^t_{-\infty} dt'\int dvol_{k'} \tilde{G}^\mu_\nu(\vc{k},t;\vc{k}',t'|\vc{Y})
\hat{F}^\nu_\rho(\vc{k},t'|\vc{Y})\,\tilde{v}^\rho(\vc{k},t'|\vc{Y})\\
&+\delta\chi\int^t_{-\infty} dt'\int dvol_{k'} \tilde{G}^\mu_\rho(\vc{k},t;\vc{k}',t'|\vc{Y})
\left(S^\rho_{\sigma;\alpha}+\Theta^\rho{}_{\sigma;\alpha}\right)(t'|\vc{Y})\,\tilde{v}^\sigma(\vc{k}',t'|\vc{Y})\\
&+\cdots.
\end{split}
\label{expanded vs 1}
\end{equation}


We assume the Gaussian-randomness on the velocity fluctuation in far past, which indicates that the bare field $\tilde{\vc{v}}$ itself is also Gaussian-random equivalently at arbitrary time. Under this assumption, arbitrary correlation of $\tilde{\vc{v}}$ is to be represented in terms of its auto-correlations. For example, the fourth-order correlation of $\tilde{\vc{v}}$ is calculated as
\begin{equation*}
\begin{split}
\langle\tilde{v}^\alpha(\vc{k}^{{}_{(1)}},t^{{}_{(1)}})&\tilde{v}^\beta(\vc{k}^{{}_{(2)}},t^{{}_{(2)}})\tilde{v}^\gamma(\vc{k}^{{}_{(3)}},t^{{}_{(3)}})\tilde{v}^\delta(\vc{k}^{{}_{(4)}},t^{{}_{(4)}})\rangle\\
=&\langle\tilde{v}^\alpha(\vc{k}^{{}_{(1)}},t^{{}_{(1)}})\tilde{v}^\beta(\vc{k}^{{}_{(2)}},t^{{}_{(2)}})\rangle\ \langle \tilde{v}^\gamma(\vc{k}^{{}_{(3)}},t^{{}_{(3)}})\tilde{v}^\delta(\vc{k}^{{}_{(4)}},t^{{}_{(4)}})\rangle\\
&+\langle\tilde{v}^\alpha(\vc{k}^{{}_{(1)}},t^{{}_{(1)}})\tilde{v}^\gamma(\vc{k}^{{}_{(3)}},t^{{}_{(3)}})\rangle\ \langle \tilde{v}^\beta(\vc{k}^{{}_{(2)}},t^{{}_{(2)}})\tilde{v}^\delta(\vc{k}^{{}_{(4)}},t^{{}_{(4)}})\rangle\\
&+\langle\tilde{v}^\alpha(\vc{k}^{{}_{(1)}},t^{{}_{(1)}})\tilde{v}^\delta(\vc{k}^{{}_{(4)}},t^{{}_{(4)}})\rangle\ \langle \tilde{v}^\beta(\vc{k}^{{}_{(2)}},t^{{}_{(2)}})\tilde{v}^\gamma(\vc{k}^{{}_{(3)}},t^{{}_{(3)}})\rangle.
\end{split}
\end{equation*}
In general, for an even positive integer $n$, the correlation of the $n$th order is reduced to  
\begin{equation*}
\begin{split}
\langle\tilde{v}^\alpha&(\vc{k}^{{}_{(1)}},t^{{}_{(1)}})\tilde{v}^\beta(\vc{k}^{{}_{(2)}},t^{{}_{(2)}})\cdots
\tilde{v}^\eta(\vc{k}^{{}_{(n-1)}},t^{{}_{(n-1)}})\tilde{v}^\zeta(\vc{k}^{{}_{(n)}},t^{{}_{(n)}})\rangle\\
=&\langle\tilde{v}^\alpha(\vc{k}^{{}_{(1)}},t^{{}_{(1)}})\tilde{v}^\beta(\vc{k}^{{}_{(2)}},t^{{}_{(2)}})\rangle\cdots
\langle\tilde{v}^\eta(\vc{k}^{{}_{(n-1)}},t^{{}_{(n-1)}})\tilde{v}^\zeta(\vc{k}^{{}_{(n)}},t^{{}_{(n)}})\rangle\\
&+\cdots\ \textrm{all the rest combinations}\ \ \ (\textrm{for even}\,n),
\end{split}
\end{equation*}
while it vanishes for an odd $n$. In the later discussions we write the bare autocorrelation $\langle\tilde{v}^\mu(\vc{k},t|\vc{Y})\tilde{v}^\nu(\vc{k}',t'|\vc{Y})\rangle$ as $\tilde{Q}^{\mu\nu}(\vc{k},t;\vc{k}',t'|\vc{Y})$. For Eq. (\ref{expanded vs 1}), an arbitrary moment of $\sol \vc{v}$ can be expressed by a series expansion of $\tilde{\mathsfbi{Q}}$ and $\tilde{\mathsfbi{G}}$ using $\lambda$, $\chi$, and $\delta$ as perturbative parameters.




Here let us mention again the arbitrariness of $\mathsfbi{A}$ introduced in Sec. \ref{ORTHONORMAL}. Since $\hat{\mathsfbi{L}}$ in Eqs. (\ref{vtilde}) and (\ref{Gtilde}) contains $\mathsfbi{A}$, both $\tilde{\vc{v}}$ and $\tilde{\mathsfbi{G}}$ depend on $\mathsfbi{A}$. Especially, for a strain-free flow, Eq. (\ref{vtilde}) becomes
\begin{equation}
\frac{\partial}{\partial t} \tilde{v}^\mu(\vc{k},t|\vc{Y})
=A^\mu{}_\rho(t|\vc{Y})\tilde{v}^\rho(\vc{k},t|\vc{Y})-\nu k^2 \tilde{v}^\mu(\vc{k},t|\vc{Y}),
\end{equation}
where the first term of the right side expresses rotation of $\tilde{\vc{v}}$; unperturbed field, whereas isotropic, has rotation relative to the mean-Lagrangian frame at angular-velocity $\mathsfbi{A}$. Then the rotation of the unperturbed field is selected by choosing $\mathsfbi{A}$. Note that the infinite series (\ref{expanded vs 1}) is invariant for an arbitrary $\mathsfbi{A}$; since Eq. (\ref{expanded eq}) is invariant for an arbitrary $\mathsfbi{A}$, the total field $\vc{v}'$ does not depend on the orthonormal frame of $SO(3)$ symmetry. In reality, however, we should truncate the series at certain finite order, and we shall choose $\mathsfbi{A}$ so that the lower-order truncation can mimic the total field as much as possible. It is well expected that the velocity fluctuation may reflect the mean-flow rotation via energy cascading from larger- to smaller scales. In this paper we choose $\tilde{\vc{v}}$ to be rotating with the mean flow, meaning no relative rotation to the mean-Lagrangian frame (i.e. $\mathsfbi{A}=\mathsfbi{0}$).


\subsection{Basic field}\label{BASIC}
Among all the terms in expansion (\ref{expanded vs 1}), there is a particular group of terms free from both $\chi$ and $\delta$. We collect them and define another vector field $\bas\vc{v}$ as their summation:
\begin{equation}
\begin{split}
\bas v^\mu&(\vc{k},t|\vc{Y})\\
=&\tilde{v}^\mu(\vc{k},t|\vc{Y})\\
&+\lambda\int^t_{-\infty} dt'\int dvol_{k'} \tilde{G}^\mu_\nu(\vc{k},t;\vc{k}',t'|\vc{Y})
\frac{1}{i}M^\nu_{\rho\sigma}[\vc{k}';\vc{p},\vc{q}]
\,\tilde{v}^\rho(\vc{p},t'|\vc{Y})
\,\tilde{v}^\sigma(\vc{q},t'|\vc{Y})\\
&+2\lambda^2\int^t_{-\infty} dt'\int dvol_{k'} \tilde{G}^\mu_\nu(\vc{k},t;\vc{k}',t'|\vc{Y})
\frac{1}{i}M^\nu_{\rho\sigma}[\vc{k}';\vc{p},\vc{q}]
\tilde{v}^\rho(\vc{p},t'|\vc{Y})\\
&\ \ \ \ \ \ \times\int^{t'}_{-\infty} dt'' \int dvol_{q'} \tilde{G}^\sigma_\kappa(\vc{q},t';\vc{q}',t''|\vc{Y})
\frac{1}{i}M^\kappa_{\alpha\beta}[\vc{q}';\vc{p}'',\vc{q}'']
\,\tilde{v}^\alpha(\vc{p}'',t''|\vc{Y})
\,\tilde{v}^\beta(\vc{q}'',t''|\vc{Y})\\
&+O(\lambda^3),
\end{split}
\label{basic field}
\end{equation}
which may be called the basic field. The basic field clearly satisfies the Navier-Stokes equation in cotangent space:
\begin{equation}
\hat{L}^\mu_\nu\, \bas v^\nu(\vc{k},t|\vc{Y})
=\lambda\frac{1}{i}M^\mu_{\rho\sigma}\cnv{k}{p}{q}
\,\bas v^\rho(\vc{p},t|\vc{Y})
\,\bas v^\sigma(\vc{q},t|\vc{Y}),
\label{bv eq}
\end{equation}
The basic field only retains pure-$\lambda$ terms, being free of both anisotropy and inhomogeneity arising from $\chi$- and $\delta$-related terms.

\subsection{Fine-Lagrangian picture}\label{FINE LAGRANGIAN}
So far, our basic quantities are described on the mean-Lagrangian coordinate, which cancels out the sweeping effect caused by the mean flow. In order to extract the timescale consistent with Kolmogorov's theory, we shall remove the sweeping effect caused by fine motion of $\bas\vc{v}$. To address this requirement, here we introduce a Lagrangian picture based on $\bas\vc{v}$. Since $\bas\vc{v}(\vc{\xi},t|\vc{Y})$ provides a vector field on a tangent space $T_y\mathscr{M}_t$, one can consider a point convected by $\bas\vc{v}(\vc{\xi},t|\vc{Y})$ within $T_y\mathscr{M}_t$. From this view point, the Lagrangian position function can be introduced as a density function of the convected point; providing the convected point locates at $\vc{\xi}'$ at time $t'$, the Lagrangian position function $\bas\psi(\vc{\xi},t;\vc{\xi}',t'|\vc{Y})$ gives the density of the convected point in $\vc{\xi}$ space. The Lagrangian position function provides a one-to-one mapping $\bas\psi$ from an arbitrary TS variable on $T\mathscr{M}_t$ (and its conjugate on $T\mathscr{M}_t$) to another TS variable, e.g. $\bas\vc{v}(\vc{\xi},t|\vc{Y})$ on $T\mathscr{M}_t$ is mapped to $\bas \vc{w}(t|\vc{\xi},t'|\vc{Y})\equiv\int d vol_{\xi''}\bas\psi(\vc{\xi}'',t;\vc{\xi},t'|\vc{Y})\bas \vc{v}(\vc{\xi}'',t|\vc{Y})$ which corresponds to the \emph{generalized velocity field} of Ref. \cite{Kaneda81} and its further extension given by Ref. \cite{Ariki17}. Then we introduce a correlation $\langle\bas w^\mu(t|\vc{\xi},t'|\vc{Y})\,\bas v^\mu(\vc{\xi}',t'|\vc{Y})\rangle$, which is invariant under the Galilean transformation $\vc{\xi}\to\vc{\xi}+\vc{a}(t-t')$ for an arbitrary constant vector $\vc{a}$, as a suitable variable representing a turbulence time scale free from sweeping effect. Another mapping, say $\bas\psi^*$, can be equally defined by the Fourier coefficients $\bas\psi(\vc{k}'',t;\vc{k},t'|\vc{Y})\equiv (2\pi)^3\mathcal{F}^{-k''}_{\xi''}\mathcal{F}^{k}_{\xi}\ \bas\psi(\vc{\xi}'',t;\vc{\xi},t'|\vc{Y})$ in the dual space $T^*\mathscr{M}_t$. Then $\bas\vc{v}(\vc{k},t|\vc{Y})$ is mapped to 
\begin{equation}
\bas w^\mu(t|\vc{k},t'|\vc{Y})
\equiv\int dvol_{k''}\bas\psi(\vc{k}'',t;\vc{k},t'|\vc{Y})\bas v^\mu(\vc{k}'',t|\vc{Y}).
\label{Lagrangian bv}
\end{equation}
Similarly, the response of $\bas \vc{w}(t|\vc{k},t'|\vc{Y})$ against infinitesimal disturbing force $\boldsymbol{\Upsilon}(\vc{k}',t'|\vc{Y})$ applied to fluid can be considered:
\begin{equation}
\bas G'^\mu_\nu(\vc{k},t;\vc{k}',t'|\vc{Y})\equiv\frac{\delta\ \bas w^\mu(t|\vc{k},t'|\vc{Y})}{\delta \Upsilon^\nu(\vc{k}',t'|\vc{Y})},
\label{Lagrangian response}
\end{equation}
where an operation $\delta/\delta\vc{\Upsilon}$ expresses the functional derivative. Here $\bas G'^\mu_\nu(\vc{k},t;\vc{k}',t'|\vc{Y})$, a tensor-valued field on $T^*\mathscr{M}_t\otimes T^*\mathscr{M}_{t'}$, represents the Lagrangian response based on $\bas\vc{v}$. Now the following two statistical quantities play the key role in the later renormalization procedure:
\begin{subequations}
\begin{align}
&\bas Q^{\mu\nu}(\vc{k},t;\vc{k}',t'|\vc{Y})
\equiv\int d vol_{k''} \langle\bas w^\mu(t|\vc{k},t'|\vc{Y})\,\bas v^\nu(\vc{k},t'|\vc{Y})\rangle,\label{BQ}\\
&\bas G^\mu_\nu(\vc{k},t;\vc{k}',t'|\vc{Y})
\equiv\langle \bas G'^\mu_\nu(\vc{k},t;\vc{k}',t'|\vc{Y})\rangle.\label{BG}
\end{align}
\end{subequations}

Perturbative expansion of $\bas\mathsfbi{Q}$ and $\bas\mathsfbi{G}$ can be performed using $\bas\psi$; the Lagrangian position function $\bas\psi(\vc{k}'',t;\vc{k},t'|\vc{Y})$ is governed by 
\begin{equation}
\frac{\partial}{\partial t} \bas\psi(\vc{k}'',t;\vc{k},t'|\vc{Y})
=i\lambda k''_\mu[\vc{k}'';-\vc{p},\vc{q}]\bas v^\mu(\vc{p},t|\vc{Y})\bas\psi(\vc{q},t;\vc{k},t'|\vc{Y}),
\label{psi eq.}
\end{equation}
with $\bas\psi(\vc{k}'',t';\vc{k},t'|\vc{Y})=\delta^3_c(\vc{k}''-\vc{k}|\vc{Y})$. Here we introduced again the nonlinearity parameter $\lambda$. Regarding the nonlinear right side of Eq. (\ref{psi eq.}) as a perturbation, we obtain an iterative expansion:
\begin{equation}
\begin{split}
\bas\psi(\vc{k}'',t;\vc{k},t'|\vc{Y})=&\tilde{\psi}(\vc{k}'',t;\vc{k},t'|\vc{Y})\\
&+\lambda\int^t_{t'} dt''\  ik''_\mu[\vc{k}'';-\vc{p},\vc{q}]\tilde{v}^\mu(\vc{p},t''|\vc{Y})\tilde{\psi}(\vc{q},t'';\vc{k},\tau|\vc{Y})\\
&+O(\lambda^2)
\end{split}
\label{psi expansion}
\end{equation}
where $\tilde{\psi}(\vc{k}'',t;\vc{k},t'|\vc{Y})\equiv\delta^3_c(\vc{k}''-\vc{k}|\vc{Y})$ is regarded as the unperturbed solution. Using Eqs. (\ref{basic field}) and (\ref{psi expansion}), $\bas\mathsfbi{Q}$ and $\bas\mathsfbi{G}$ can be expanded by $\tilde{\mathsfbi{Q}}$ and $\tilde{\mathsfbi{G}}$.

\subsection{Renormalization on $\lambda$}\label{RENORMALIZATION}
Up to now, we have setup a systematic procedure of perturbation analysis; all the physical quantities of our interest are to be expanded in terms of $\tilde{\mathsfbi{Q}}$ and $\tilde{\mathsfbi{G}}$ with the perturbative parameters $\lambda$, $\chi$, and $\delta$. While $\chi$ and $\delta$ may be regarded as small parameters at certain physical conditions, we can hardly treat $\lambda$-related terms as small perturbations in case of fully-developed turbulence, so we need to \emph{renormalize} these terms. What follows is to utilize these $\bas \mathsfbi{Q}$ and $\bas\mathsfbi{G}$ to incorporate the strong nonlinearity into the simple perturbation analysis; we replace $\tilde{\mathsfbi{Q}}$ and $\tilde{\mathsfbi{G}}$ with alternative expansion basis $\bas \mathsfbi{Q}$ and $\bas\mathsfbi{G}$, so that the nonlinearlity is incorporated in this series expansion even at zeroth order of $\lambda$. This is a rough sketch of the renormalized perturbation theory of Ref. \cite{Kraichnan77,Kaneda81} (also see \cite{renormalization}). The later procedure may be summarized as follows:
\begin{enumerate}
\item Due to the Gaussian nature of the bare field, arbitrary moment $F$ of our concern can be expanded in terms of $\lambda$, $\chi$ and $\delta$ and is represented as a series expansion $F=\sum_{l,m,n=0}^\infty \lambda^l \chi^m\delta^n F_{lmn}[\tilde{\mathsfbi{Q}},\tilde{\mathsfbi{G}}]$, where $F_{lmn}[\cdots,\cdots]$ is a functional.\\
\item On the other hand, the basic quantities $\bas \mathsfbi{Q}$ and $\bas\mathsfbi{G}$ are to be expanded only by $\lambda$ and are represented as series expansions: $\bas \mathsfbi{Q}=\sum_{l=0}^\infty\lambda^l\mathsfbi{Q}_l[\tilde{\mathsfbi{Q}},\tilde{\mathsfbi{G}}]$ and $\bas \mathsfbi{G}=\sum_{l=0}^\infty\lambda^l\mathsfbi{G}_l[\tilde{\mathsfbi{Q}},\tilde{\mathsfbi{G}}]$.\\
\item  Then \emph{invert} these expansions and obtain the series expansions by the basic quantities: $\tilde{\mathsfbi{Q}}=\sum_{l=0}^\infty\lambda^l\tilde{\mathsfbi{Q}}_l[\bas\mathsfbi{Q},\bas\mathsfbi{G}]$ and $\tilde{\mathsfbi{G}}=\sum_{l=0}^\infty\lambda^l\tilde{\mathsfbi{G}}_l[\bas\mathsfbi{Q},\bas\mathsfbi{G}]$.\\
\item By substituting these inverted expansions into $F=\sum_{l,m,n=0}^\infty \lambda^l \chi^m\delta^n F_{lmn}[\tilde{\mathsfbi{Q}},\tilde{\mathsfbi{G}}]$, we obtain another series expansion $F$ $=\sum_{l,m,n=0}^\infty$ $\lambda^l \chi^m\delta^n \mathscr{F}_{lmn}[\bas\mathsfbi{Q},\bas\mathsfbi{G}]$. By taking non-zero lowest-order truncation with respect to $\lambda$, we obtain the demanded approximation of $F$.
\end{enumerate}
Providing the lowest order is $l=L$, we obtain
\begin{equation}
F\overset{\!\!\scalebox{0.4}{LRA}}{\longrightarrow}\lambda^L \sum_{m,n=0}^\infty \chi^m\delta^n \mathscr{F}_{Lmn}[\bas\mathsfbi{Q},\bas\mathsfbi{G}].
\label{re F}
\end{equation}
Now the nonlinear interaction is incorporated in this series expansion even at the lowest order of $\lambda$. The above set of procedures 1-4 forms the core idea of \textit{Lagrangian renormalization approximation} (LRA) of Ref. \cite{Kaneda81}. The key factors $\bas \mathsfbi{Q}$ and $\bas\mathsfbi{G}$ are referred to as the renormalized correlation and propagator respectively. The above procedure can also be utilized to obtain the dynamical equations of $\bas\mathsfbi{Q}$ and $\bas\mathsfbi{G}$, so arbitrary moments of $\bas \vc{v}$, in principle, can be calculated in this procedure \cite{intm}. Using the orthonormal-frame representations, i.e. $\bas Q^{IJ}(\check{\vc{k}},t;\check{\vc{k}}',t')$ and $\bas G^I_J(\check{\vc{k}},t;\check{\vc{k}}',t')$, we write
\begin{subequations}
\begin{align}
&\bas Q^{\mu\nu}(\vc{k};t,\vc{k}',t')
=e^\mu_I(t)e^\nu_J(t')\bas Q^{IJ}(\bar{\mathsfbi{e}}(t)\vc{k},t;\bar{\mathsfbi{e}}(t')\vc{k}',t'),\\
&\bas G^\mu_\nu(\vc{k};t,\vc{k}',t')
=e^\mu_I(t)\bar{e}_\nu^J(t')\bas G^I_J(\bar{\mathsfbi{e}}(t)\vc{k},t;\bar{\mathsfbi{e}}(t')\vc{k}',t').
\end{align}
\label{isotropic QG0}
\end{subequations}
For homogeneity and isotropy, the orthonormal-frame representations of $\bas\mathsfbi{Q}$ and $\bas\mathsfbi{G}$ are reduced to
\begin{subequations}
\begin{align}
\bas Q^{IJ}(\check{\vc{k}},t;\check{\vc{k}}',t')&=\frac{1}{2}P^{IJ}(\check{\vc{k}})\ 
\bas Q(\check{k};t,t')\delta^3(\check{\vc{k}}+\check{\vc{k}}'),\\
\bas G^I_J(\check{\vc{k}},t;\check{\vc{k}}',t')&=P^I_J(\check{\vc{k}})\ 
\bas G(\check{k};t,t')\delta^3(\check{\vc{k}}-\check{\vc{k}}'),
\end{align}
\label{isotropic QG}
\end{subequations}
where $\bas Q$ and $\bas G$ are both scalar functions of the orthonormal-wavenumber radius $\check{k}\equiv (\delta^{IJ}\check{k}_I \check{k}_J)^{1/2}$. Applying the renormalization procedure, we reach a closed set of equations for $\bas Q$ and $\bas G$
\begin{equation}
\begin{split}
&\left(\frac{\partial}{\partial t}+2\nu \check{k}^2\right)\bas Q(\check{k},t,t|\vc{Y})\\
&=2\pi\iint_\triangle d\check{p}\,d\check{q}\,\check{p}^2 \check{q}(xy+z^3)\int^t_{-\infty} ds\\
&\ \ \ \ \ \times \bas Q(\check{q};t,t'|\vc{Y})
\Big\{\bas Q(\check{p};t,s|\vc{Y})\bas G(\check{k};t,s|\vc{Y})-\bas Q(\check{k};t,s|\vc{Y}) \bas G(\check{p};t,s|\vc{Y})\Big\},
\end{split}
\label{E eq}
\end{equation}
\begin{equation}
\begin{split}
&\left(\frac{\partial}{\partial t}+\nu \check{k}^2\right)\bas G(\check{k},t,t'|\vc{Y})\\
&=-\pi\iint_\triangle d\check{p}\,d\check{q}\,\check{k}\check{p}\check{q}(1-y^2)(1-z^2)\int^t_{t'} ds\bas Q(\check{q};t,s|\vc{Y})\bas G(\check{k},t,t'|\vc{Y}),
\end{split}
\label{G eq}
\end{equation}
\begin{equation}
\bas Q(\check{k},t,t'|\vc{Y})=\bas Q(\check{k},t',t'|\vc{Y})\bas G(\check{k},t,t'|\vc{Y}),
\label{FD}
\end{equation}
which are identical to Eqs. (2.48)-(2.53) of Ref. \cite{Kaneda81} except for the dependence on $\vc{Y}$ ($x\equiv(\check{p}^2+\check{q}^2-\check{k}^2)/(2\check{p}\check{q})$ $y\equiv(\check{q}^2+\check{k}^2-\check{p}^2)/(2\check{k}\check{q})$, $z\equiv(\check{p}^2+\check{k}^2-\check{q}^2)/(2\check{k}\check{p})$, and $\triangle\equiv\left\{(\check{p},\check{q})\right|\left.|\check{k}-\check{p}|\leq \check{q} \leq \check{k}+\check{p}\right\}$). In our formalism, we calculate an arbitrary moment $F$ from Eqs. (\ref{re F})-(\ref{FD}). 

In concluding this chapter, we shall mention the relationship between TSLRA and LRA. The obtained closure scheme, which is finally attributed to Eq. (\ref{re F}), exactly coincides with LRA at the order of $O(\chi^0\delta^0)$, implying that TSLRA is an extension of LRA subjected to anisotropy and inhomogeneity. Then, remark that the above renormalization retains only the contributions from $\lambda$, and we merely performed simple perturbative expansions for $\chi$ and $\delta$. In this sense our renormalization procedure remains \textit{partial}, regarding further renormalization on $\chi$- and $\delta$ effects may be possible to incorporate strong anisotropy and inhomogeneity. Such attempts may leads to so called the \emph{second-order closure}, where effective turbulence fluxes such as the Reynolds stress are to be treated as independent dynamical variables. 

\section{Application to the Reynolds stress}\label{REYNOLDS STRESS}
\subsection{space-time non-local expression}\label{space-time non-local expression}
\begin{figure}
\centering
\includegraphics[width=13cm]{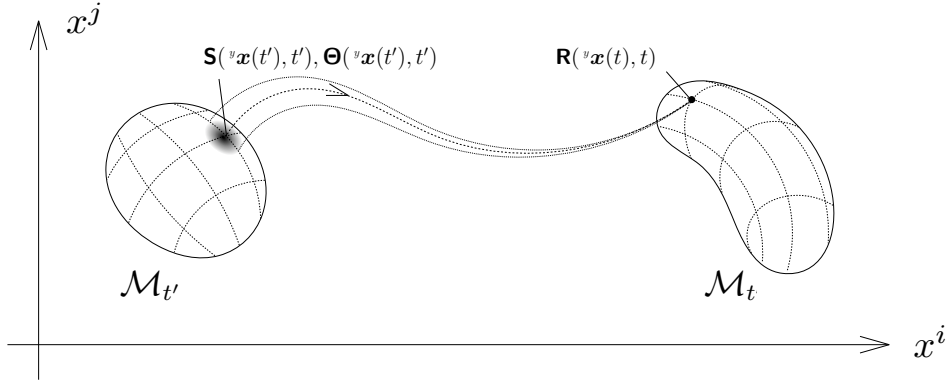}
\caption{The Reynolds stress at ${}^{{}_y}\vc{x}(t)\in\mathcal{M}_t$ incorporates the memories of $\mathsfbi{S}$ and $\mathsfbi{\Theta}$ around the trajectory ${}^{{}_y}\vc{x}(t')$.}
\label{memory}
\end{figure}

As a typical example, let us calculate the Reynolds stress $R^{\mu\nu}\equiv\langle v'^\mu v'^\nu\rangle$ in our formalism. In the wavenumber-space (cotangent-space) representation, this corresponds to $\langle v'^\mu(\vc{k},t|\vc{Y})v'^\nu(\vc{k}',t|\vc{Y})\rangle$, which is a tensor-valued function on $T^*\mathscr{M}_t\otimes T^*\mathscr{M}_t$. By integrating over $T^*_y\mathscr{M}_t\otimes T^*_y\mathscr{M}_t$, this is projected onto $\mathscr{M}_t$, yielding an observable field on $\mathscr{M}_t$:
\begin{equation}
\begin{split}
R^{\mu\nu}(\vc{y},t)=\left.\iint dvol_{k}\,dvol_{k'} \langle v'^\mu(\vc{k},t|\vc{Y})v'^\nu(\vc{k}',t|\vc{Y})\rangle\right|_{\vc{Y}=\delta\vc{y}}.
\end{split}
\label{spectrum represented R}
\end{equation}
As previously discussed, we apply the renormalized perturbation procedures (1-4 in Sec. \ref{RENORMALIZATION}) to a correlation $\langle v'^\mu(\vc{k},t|\vc{Y})v'^\nu(\vc{k}',t|\vc{Y})\rangle$ to obtain its series expansion in the form of Eq. (\ref{re F}). Let us first demonstrate the calculation of the lowest-order term of Eq. (\ref{spectrum represented R}). Using the orthonormal representation, we obtain
\begin{equation*}
\begin{split}
&\iint dvol_{k}\,dvol_{k'} \langle \bas v^\mu(\vc{k},t|\vc{Y})\,\bas v^\nu(\vc{k}',t|\vc{Y})\rangle\\
&=\iint dvol_{k}\,dvol_{k'} \bas Q^{\mu\nu}(\vc{k},t;\vc{k}',t|\vc{Y})\\
&=e^\mu_I(t|\vc{Y})e^\mu_J(t|\vc{Y})\iint d^3\check{k}\,d^3\check{k}' \ \bas Q^{IJ}(\check{\vc{k}},t;\check{\vc{k}}',t|\vc{Y})\\
&=e^\mu_I(t|\vc{Y})e^\mu_J(t|\vc{Y})\int d^3\check{k}\int d^3\check{k}'
\frac{1}{2}P^{IJ}(\check{k}|\vc{Y})
\ \bas Q(\check{k};t,t|\vc{Y})\delta^3(\check{\vc{k}}+\check{\vc{k}}'|\vc{Y})\\
&=e^\mu_I(t|\vc{Y})e^\mu_J(t|\vc{Y})\frac{1}{3}\delta^{IJ}\int d^3\check{k}\ \bas Q(\check{k};t,t|\vc{Y})\\
&=\frac{1}{3}g^{\mu\nu}(t|\vc{Y})\int_0^\infty 4\pi\check{k}^2 d\check{k}\ \bas Q(\check{k};t,t|\vc{Y})\\
&\rightarrow \frac{1}{3}g^{\mu\nu}(\vc{y},t)\int_0^\infty 4\pi\check{k}^2 d\check{k}\ 
 \bas Q(\check{k};t,t|\vc{y})\  (\vc{Y}\rightarrow \delta \vc{y},\ \delta\rightarrow 1),
\end{split}
\end{equation*}
where we used a formula 
\begin{equation*}
\int d^3\check{k}\ \frac{\check{k}_I\check{k}_J}{\check{k}^2} F_\mathrm{iso}(\check{k})
=\frac{1}{3}\delta_{IJ}\int^\infty_0 4\pi\check{k}^2 d\check{k}\, F_\mathrm{iso}(\check{k})
\label{2nd int}
\end{equation*}
for an arbitrary isotropic function $F_\mathrm{iso}(\check{k})$. For higher-order analyses, this may be generalized as
\begin{equation*}
\int d^3\check{k}\ \frac{\check{k}_I\check{k}_J\cdots\check{k}_K\check{k}_L}{\check{k}^2} F_\mathrm{iso}(\check{k})
=\frac{1}{\Pi_{m=1}^n (2m+1)}\delta^{\scalebox{0.7}{(2n)}}_{IJ\cdots KL} \int^\infty_0 4\pi\check{k}^2 d\check{k}\, F_\mathrm{iso}(\check{k}) \ \ (n\in\mathbb N),
\label{nth int}
\end{equation*}
where $\vc{\delta}^{\scalebox{0.7}{(2n)}}$ is $2n$-rank isotropic tensor, e.g. the four-rank ($n=2$) isotropic tensor reads $\delta^{\scalebox{0.7}{(4)}}_{IJKL}
=\delta_{IJ}\delta_{KL}+\delta_{IK}\delta_{JL}+\delta_{IL}\delta_{JK}$. Likewise, the orthonormal frame offers a basic representation to integrate isotropic functions. Following the similar steps, substituting the renormalized perturbation expansion of $\langle v'^\mu(\vc{k},t|\vc{Y})v'^\nu(\vc{k}',t|\vc{Y})\rangle$ into Eq. (\ref{spectrum represented R}) yields
\begin{equation}
\begin{split}
R^{\mu\nu}(\vc{y}&,t)\\
=&\frac{1}{3}\int_0^\infty 4\pi\check{k}^2 d\check{k}\ \bas Q(\check{k};t,t|\vc{y}) g^{\mu\nu}(\vc{y},t)\\
&-\frac{7}{30}\chi\int^t_{-\infty} dt'\ \left\{
\Lambda^\mu_\rho(\vc{y};t,t')\bar{\Lambda}^\sigma_\kappa(\vc{y};t,t')g^{\kappa\nu}(\vc{y},t)
+\Lambda^\nu_\rho(\vc{y};t,t')\bar{\Lambda}^\sigma_\kappa(\vc{y};t,t')g^{\kappa\mu}(\vc{y},t)
\right\}\\
&\ \ \ \ \ \ \ \ \times\left(\frac{1}{2}S^\rho_\sigma+\Theta^\rho{}_{\sigma}\right)(\vc{y},t')\int_0^\infty 4\pi\check{k}^2 d\check{k} \ \bas G(\check{k};t,t'|\vc{y})\bas Q(k;t,t'|\vc{y})\\
&-\frac{1}{20}\chi\int^t_{-\infty} dt'\ \left\{
g^{\mu\rho}(\vc{y},t)S^\nu_\rho(\vc{y},t')+g^{\nu\rho}(\vc{y},t)S^\mu_\rho(\vc{y},t')
\right\}\\
&\ \ \ \ \ \ \ \ \times\int_0^\infty 4\pi\check{k}^2 d\check{k}\ \left\{
\bas G(\check{k};t,t'|\vc{y})\bas Q(\check{k};t,t'|\vc{y})
+\frac{2}{3}\bas G(\check{k};t,t'|\vc{y})\frac{\partial\ \bas Q}{\partial \check{k}}(\check{k};t,t'|\vc{y})
\right\}\\
&+O(\chi^2,\delta^2),
\end{split}
\label{nonlocal}
\end{equation}
where up to $\chi$-first-order terms are written explicitly. We should remark that all the integrals in Eq. (\ref{nonlocal}) at $\vc{y}$ on $\mathscr{M}_t$ correspond to the mean-convective integration \cite{Ariki17} along the trajectory $\mlpind\vc{x}(t')$ in the physical space $\{x^{\scalebox{0.6}{1}},x^{\scalebox{0.6}{2}},x^{\scalebox{0.6}{3}}\}$. In addition, neighborhood of $y$ can also contributes to the integration regarding $\delta$-related series as a derivative expansion. Thus, in general coordinate representation, Eq. (\ref{nonlocal}) integrates the memories of $\mathsfbi{S}$ and $\mathsfbi{\Theta}$ on the nonlocal trajectory around $\mlpind\vc{x}(t')$ in the coordinate space in generally covariant manner (see Fig \ref{memory}), incorporating $\bas Q$ and $\bas G$ as the memory functions. As remarked in Ref. \cite{Ariki17}, such convective integration accounts the true memories of turbulence properties along with the fluid's path and deformation, which enables physically objective (more precisely, covariant) treatments of the history effect. It is also worth comparing Eq. (\ref{nonlocal}) with the history effects given by Refs. \cite{Hamba17,HD08}. As their remarks, history effect may appears as a natural consequence from the transport equation of the Reynolds stress, expressing the relaxation time by single-time statistics alone. In the present formalism, on the other hand, the memory functions arise from two-time statistics; TSLRA, as well as TSDIA, essentially incorporates two-time statistics, bridging the gap between homogeneous and inhomogeneous turbulence closures. 

\subsection{Space-time local expression}
While Eq. (\ref{nonlocal}) expresses the history effect of turbulence with enough clarity, many things are to be understood from space-time local expression owing to its mathematical simplicity. To see contributions from the current space-time points, we expand $\mathsfbi{S}(\vc{y},t')$, $\mathsfbi{\Theta}(\vc{y},t')$, $\mathsfbi{\Lambda}(\vc{y};t',t)$, and $\bar{\mathsfbi{\Lambda}}(\vc{y};t',t)$ around $t$ in Eq. (\ref{nonlocal}):
\begin{equation}
\begin{split}
S^\rho_\sigma(\vc{y},t')&=S^\rho_\sigma(\vc{y},t)+(t'-t)\frac{\partial S^\rho_\sigma}{\partial t}(\vc{y},t)+\frac{1}{2}(t'-t)^2\frac{\partial^2 S^\rho_\sigma}{\partial t^2}(\vc{y},t)+\cdots\ ,\\
\Theta^\rho{}_\sigma(\vc{y},t')&=\Theta^\rho{}_\sigma(\vc{y},t)+(t'-t)\frac{\partial \Theta^\rho{}_\sigma}{\partial t}(\vc{y},t)+\frac{1}{2}(t'-t)^2\frac{\partial^2 \Theta^\rho{}_\sigma}{\partial t^2}(\vc{y},t)+\cdots\ ,\\
\Lambda^\rho_\sigma(\vc{y};t',t)&=\delta^\rho_\sigma-\frac{1}{2}(t'-t)S^\rho_\sigma(\vc{y},t)-\frac{1}{4}(t'-t)^2\left(S^\rho_\alpha S^\alpha_\sigma+\frac{\partial S^\rho_\sigma}{\partial t}\right)(\vc{y},t)+\cdots\ ,\\
\bar{\Lambda}^\rho_\sigma(\vc{y}|t';t)&=\delta^\rho_\sigma+\frac{1}{2}(t'-t)S^\rho_\sigma(\vc{y},t)-\frac{1}{4}(t'-t)^2\left(S^\rho_\alpha S^\alpha_\sigma+\frac{\partial S^\rho_\sigma}{\partial t}\right)(\vc{y},t)+\cdots\ .\\
\end{split}
\label{local approx}
\end{equation}
By substituting Eqs. (\ref{local approx}) into Eq. (\ref{nonlocal}) expanded up to $O(\chi^2,\delta^2)$, $\mathsfbi{S}$ and $\mathsfbi{\Theta}$ can be split out from the memory integration, leading to a space-time local expression of Eq. (\ref{nonlocal}):
\begin{equation}
\begin{split}
R^{\mu\nu}
=&\frac{2}{3}\,\bask g^{\mu\nu}
-\chi\nu_TS^{\mu\nu}
+\chi\gamma_t\left(\frac{\partial S^{\mu\nu}}{\partial t} + S^\mu_\rho S^{\nu\rho}\right)\\
&+\chi^2N_I\ \boldsymbol{\mathsf{S\cdot S}}g^{\mu\nu}
+\chi^2N_{I\!\!I}\ \boldsymbol{\mathsf{\Theta\cdot \Theta}}g^{\mu\nu}\\
&+\chi^2N_{I\!\!I\!\!I}\ S^\mu_\rho S^{\nu\rho}
+\chi^2N_{I\!V}\ \Theta^\mu{}_\rho \Theta^{\nu\rho}\\
&+\chi^2N_{V} \left(S^\mu_\rho \Theta^{\nu\rho}+S^\nu_\rho \Theta^{\mu\rho}\right)\\
&+\delta^2\chi D_I \left(S^\mu_\rho{}^{;\nu\rho}+S^\nu_\rho{}^{;\mu\rho}\right)\\
&+\delta^2\chi D_{I\!\!I}\, \left(\Theta^{\mu}{}_{\rho}{}^{;\nu\rho}+\Theta^{\nu}{}_{\rho}{}^{;\mu\rho}\right)\\
&+\delta^2\chi D_{I\!\!I\!\!I}\ \Delta S^{\mu\nu},
\end{split}
\label{R0}
\end{equation} 
where up to second-order terms in $\boldsymbol{\mathsf{S}}$ and $\boldsymbol{\mathsf{\Theta}}$ are retained. $\bas K$, $\nu_T$, $\gamma_t$, $N_I$-$N_V$ and $D_I$-$D_{I\!V}$ are all scalars explained by the fluctuation properties as follows:
\begin{equation}
\bask=\frac{1}{2}\int_0^\infty 4\pi\check{k}^2d\check{k}\ \bas Q(\check{k};t,t),
\label{KB}
\end{equation}

\begin{equation}
\begin{split}
\nu_T=\int_0^\infty 4\pi\check{k}^2d\check{k}\int^{t}_{-\infty}dt' 
\bigg[\frac{1}{3}\bas G(\check{k};t,t')\,\bas Q(\check{k};t,t')
+\frac{1}{15}\bas G(\check{k};t,t')\check{k}\frac{\partial\bas Q}{\partial \check{k}}(\check{k};t,t')\bigg],
\end{split}
\label{nuT}
\end{equation}

\begin{equation}
\begin{split}
\gamma_t=\int_0^\infty \!\!4\pi\check{k}^2d\check{k}
\int^{t}_{-\infty}\!\!dt'(t-t') 
\bigg[\frac{1}{3}\bas G(\check{k};t,t')\bas Q(\check{k};t,t')
+\frac{1}{15}\bas G(\check{k};t,t')\check{k}\frac{\partial\bas Q}{\partial \check{k}}(\check{k};t,t')\bigg],
\end{split}
\label{gammat}
\end{equation}

\begin{equation}
\begin{split}
N_I=\int_0^\infty 4\pi\check{k}^2d\check{k}& \int^{t}_{-\infty}dt'\ \int^{t'}_{-\infty}dt''\\
\times\bigg[&\frac{1}{35}\bas G(\check{k};t,t')\,\bas G(\check{k};t',t'')\,\bas Q(\check{k};t,t'')\\
&+\frac{13}{105}\check{k}\frac{\partial \bas G}{\partial \check{k}}(\check{k};t,t')\,\bas G(\check{k};t',t'')\bas Q(\check{k};t,t'')\\
&+\frac{1}{105}\bas G(\check{k};t,t')\,\check{k}\frac{\partial \bas G}{\partial \check{k}}(\check{k};t',t'')\,\bas Q(\check{k};t,t'')\\
&+\frac{1}{35}\check{k}^2\frac{\partial}{\partial \check{k}}\left\{\frac{\partial\, \bas G}{\partial \check{k}}(k;t,t')\,\bas G(\check{k};t',t'')\right\}\bas Q(\check{k};t,t'')\bigg]\\
+\int_0^\infty 4\pi\check{k}^2d\check{k}& \int^{t}_{-\infty}dt'\ \int^t_{-\infty}dt''\\
\times\bigg[&\frac{1}{42}\bas G(\check{k};t,t')\,\bas G(\check{k};t,t'')\,\bas Q(\check{k};t',t'')\\
&+\frac{1}{105}\check{k}\frac{\partial \bas G}{\partial \check{k}}(\check{k};t,t')\,\bas G(\check{k};t,t'')\,\bas Q(\check{k};t',t'')\\
&+\frac{1}{70}\check{k}^2\frac{\partial \bas G}{\partial \check{k}}(\check{k};t,t')\,\frac{\partial \bas G}{\partial \check{k}}(\check{k};t,t'')\,\bas Q(\check{k};t',t'')\bigg],
\end{split}
\label{NI}
\end{equation}

\begin{equation}
\begin{split}
N_{I\!\!I}=&-\frac{1}{15}\int_0^\infty 4\pi\check{k}^2d\check{k} \int^{t}_{-\infty}dt'
\int^{t'}_{-\infty}dt''
\bas G(\check{k};t,t')\,\bas G(\check{k};t',t'')\,\bas Q(\check{k};t,t'')\\
&+\frac{1}{30}\int_0^\infty 4\pi\check{k}^2d\check{k} \int^{t}_{-\infty}dt'\int^t_{-\infty}dt''
\bas G(\check{k};t,t')\,\bas G(\check{k};t,t'')\,\bas Q(\check{k};t',t''),
\end{split}
\label{NII}
\end{equation}

\begin{equation}
\begin{split}
N_{I\!\!I\!\!I}=\int_0^\infty 4\pi\check{k}^2d\check{k}& \int^{t}_{-\infty}dt'\ \int^{t'}_{-\infty}dt''\\
\times\bigg[&-\frac{5}{42}\bas G(\check{k};t,t')\,\bas G(\check{k};t',t'')\,\bas Q(\check{k};t,t'')\\
&-\frac{11}{70}\check{k}\frac{\partial \bas G}{\partial \check{k}}(\check{k};t,t')\,\bas G(\check{k};t',t'')\bas Q(\check{k};t,t'')\\
&-\frac{13}{210}\bas G(\check{k};t,t')\,\check{k}\frac{\partial \bas G}{\partial \check{k}}(\check{k};t',t'')\,\bas Q(\check{k};t,t'')\\
&-\frac{1}{210}\check{k}^2\frac{\partial^2\, \bas G}{\partial \check{k}^2}(\check{k};t,t')\,\bas G(\check{k};t',t'')\,\bas Q(\check{k};t,t'')\bigg]\\
+\int_0^\infty 4\pi\check{k}^2d\check{k}& \int^{t}_{-\infty}dt'\ \int^t_{-\infty}dt''\\
\times\bigg[&\frac{1}{21}\bas G(\check{k};t,t')\,\bas G(\check{k};t,t'')\,\bas Q(\check{k};t',t'')\\
&-\frac{13}{210}\check{k}\frac{\partial \bas G}{\partial \check{k}}(\check{k};t,t')\,\bas G(\check{k};t,t'')\,\bas Q(\check{k};t',t'')\\
&+\frac{1}{105}\check{k}^2\frac{\partial \bas G}{\partial \check{k}}(\check{k};t,t')\,\frac{\partial \bas G}{\partial \check{k}}(\check{k};t,t'')\,\bas Q(\check{k};t',t'')\bigg],
\end{split}
\label{NIII}
\end{equation}

\begin{equation}
\begin{split}
N_{I\!V}=\int_0^\infty 4\pi\check{k}^2d\check{k}
\int^{t}_{-\infty}dt'\int^{t'}_{-\infty}dt''\bigg[&-\frac{2}{15}\bas G(\check{k};t,t')\,\bas G(\check{k};t',t'')\,\bas Q(\check{k};t,t'')\\
&+\frac{1}{15}\bas G(\check{k};t,t')\,\bas G(\check{k};t,t'')\,\bas Q(\check{k};t',t'')\bigg],
\end{split}
\label{NIV}
\end{equation}

\begin{equation}
\begin{split}
N_V=\int_0^\infty 4\pi\check{k}^2d\check{k} \int^{t}_{-\infty}dt'\ \int^{t'}_{-\infty}dt''
\bigg[&\frac{1}{30}\bas G(\check{k};t,t')\,\bas G(\check{k};t',t'')\,\bas Q(\check{k};t,t'')\\
&+\frac{1}{20}\ \bas G(\check{k};t,t')\,\bas G(\check{k};t,t'')\,\bas Q(\check{k};t',t'')\bigg],
\end{split}
\label{NV}
\end{equation}

\begin{equation}
D_I=\frac{1}{14}\int_0^\infty 4\pi\check{k}^2d\check{k} \int^t_{-\infty}dt' \check{k}^{-2}\,\bas G(\check{k};t,t')\, \bas Q(\check{k};t,t'),
\label{DI}
\end{equation}


\begin{equation}
D_{I\!I}=-\frac{1}{30}\int_0^\infty 4\pi\check{k}^2d\check{k} \int^t_{-\infty}dt' k^{-2}\,\bas G(\check{k};t,t')\, \bas Q(\check{k};t,t'),
\label{DII}
\end{equation}

\begin{equation}
D_{I\!\!I\!\!I}=-\frac{1}{21}\int_0^\infty 4\pi\check{k}^2d\check{k} \int^t_{-\infty}dt' \check{k}^{-2}\,\bas G(\check{k};t,t')\, \bas Q(\check{k};t,t').
\label{DIII}
\end{equation}
In the general coordinate system $\{\mathbf{x}\}$, Eq. (\ref{R0}) turns into
\begin{equation}
\begin{split}
R^{ij}=&\frac{\partial x^i}{\partial y^\mu}\frac{\partial x^j}{\partial y^\nu}R^{\mu\nu}\\
=&\frac{2}{3}\,\bask g^{ij}
-\chi \nu_T S^{ij}
+\chi\gamma_t\left(\frac{\mathfrak{D} S^{ij}}{\mathfrak{D} t} + S^i_a S^{ja}\right)\\
&+\chi^2N_I\ \boldsymbol{\mathsf{S\cdot S}}g^{ij}
+\chi^2N_{I\!\!I}\ \boldsymbol{\mathsf{\Theta\cdot \Theta}}g^{ij}\\
&+\chi^2N_{I\!\!I\!\!I}\ S^i_a S^{ja}
+\chi^2N_{I\!V}\ \Theta^i{}_a\Theta^{ia}\\
&+\chi^2N_{V} \left(S^i_a \Theta^{ja}+S^j_a \Theta^{ia}\right)\\
&+\delta^2\chi D_I \left(S^i_a{}^{;ja}+S^j_a{}^{;ia}\right)\\
&+\delta^2\chi D_{I\!I}\, \left(\Theta^{i}{}_{a}{}^{;ja}+\Theta^{j}{}_{a}{}^{;ia}\right)\\
&+\delta^2\chi D_{I\!\!I\!\!I}\ \Delta S^{ij}.
\end{split}
\label{R1}
\end{equation} 
The third term of the right side arises from the derivative expansion of Eqs. (\ref{local approx}), representing the delay response of the stress on the strain rate. Due to generally-covariant formulation of the history effect, the delay response is now given by the convective derivative of the strain rate, expressing a physically objective property of mean-flow deformation. This may be contrasted with the Lagrangian derivative $D\mathsfbi{S}/Dt$ in TSDIA results e.g. Eq. (\ref{TSDIA}) (also see Eq. (74) in \cite{Yoshizawa84}). 

The present analysis includes $\chi^1$-, $\chi^2$- and $\chi\delta^2$-order terms. $\chi^2\delta^2$-order analysis, which produces an enormous number of terms, will be left for future studies. $\delta^1$-order terms, which includes odd numbers of wavenumber vectors, are canceled out by wavenumber integration. For later discussions, we turn back the bookkeeping parameters: $\lambda$, $\chi$, $\delta$ $\rightarrow$ 1.\\

The eddy viscosity $\nu_T$ of Eq. (\ref{nuT}) naturally extends the known interscale interaction of the homogeneous turbulence. On the right side of Eq. (\ref{E eq}), one can extract the contribution from higher-wavenumber modes $\bas Q(\check{p})$ of $\check{p}\geq\check{k}_0\gg \check{k}$:
\begin{equation}
\left(\frac{\partial}{\partial t}+2 \nu \check{k}^2\right)\,\bas Q(\check{k};t,t)
=-2\nu_T(\check{k}_0)\check{k}^2\,\bas Q(\check{k};t,t)+(\textrm{contribution from}\ \check{p}\lesssim \check{k}_0),
\end{equation}
where
\begin{equation}
\!\nu_T(\check{k}_0)=\frac{2\pi}{15}
\int_{\check{k}_0}^\infty\!\!\check{p}^2d\check{p}\int^{t}_{-\infty}\!\!dt' 
\bigg[7\bas G(\check{p};t,t')\,\bas Q(\check{p};t,t')
+\bas G^2(\check{p};t,t')\check{p}\frac{\partial\bas Q}{\partial \check{p}}(\check{p};t',t')\bigg],
\label{nuT LRA}
\end{equation}
which is identical to what is obtained by LRA and ALHDIA (see Eqs. (19)-(22) in Ref. \cite{Kaneda86} and Eqs. (3.15)-(3.16) in Ref. \cite{Kraichnan66}). On the other hand, applying partial integration to the second term of Eq. (\ref{nuT}) yields
\begin{equation}
\nu_T=\frac{2\pi}{15}\int_0^\infty \check{k}^2d\check{k}\int^{t}_{-\infty}dt' 
\bigg[7\bas G(\check{k};t,t')\,\bas Q(\check{k};t,t')
+\bas G^2(\check{k};t,t')\check{k}\frac{\partial\bas Q}{\partial \check{k}}(\check{k};t',t')\bigg],
\label{nuT TSLRA}
\end{equation}
which agrees with Eq. (\ref{nuT LRA}) for $\check{k}_0\to 0$. The coincidence between Eqs. (\ref{nuT LRA}) and (\ref{nuT TSLRA}) indicates that the scale-non-local coupling in the homogeneous-isotropic turbulence is reproduced as the lowest-order contribution of the mean-fluctuation coupling, while the higher-order properties --- $\gamma_T$, $N_I$, $N_{I\!I}$, $N_{I\!\!I\!\!I}$, $N_{I\!V}$, $N_{V}$, $D_I$, $D_{I\!I}$, and $D_{I\!\!I\!\!I}$ --- give a departure from LRA; TSLRA naturally extends the scale-non-local coupling of homogeneous isotropic turbulence to the mean-fluctuation coupling of inhomogeneous turbulence. In contrast, TSDIA, although based on a similar steps, results in
\begin{equation}
\nu_{T\ \scalebox{0.6}{TSDIA}}=\frac{7}{15}\int_0^\infty 2\pi\check{k}^2d\check{k}\int^{t}_{-\infty}dt' 
\bas G(\check{k};t,t')\,\bas Q(\check{k};t,t'),
\label{nuT TSDIA}
\end{equation}
which lacks the second integrand of Eqs. (\ref{nuT LRA}) and (\ref{nuT TSLRA}). This is exactly comes from the fact that TSDIA, for its Eulerian formulation, applies the TS representation to the mean velocity field in the convection term ($\partial/\partial t+V^j\partial/\partial x^j$); the mean velocity $\vc{V}$ may act as a uniform vector field in $\vc{\xi}$ space, improperly neglecting the mean-velocity gradient acting on the velocity fluctuation. In TSLRA, on the contrary, such non-uniformity of the mean flow is considered as a distortion of the mean-Lagrangian frame via the convective derivative $\mathfrak{D}/\mathfrak{D}t$ in Eq. (\ref{v' eq.}) ($\partial/\partial t$ in Eq. (\ref{v' eq. 2}) ) which survives TS representation.

\subsection{Inertial-range analysis}\label{INERTIAL RANGE}
\begin{figure}
\centering
\includegraphics[width=8cm]{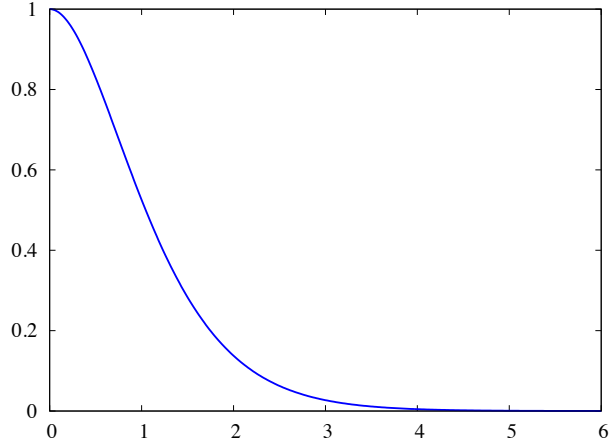}
\caption{Configuration of the dimensionless function $g(\tau)$. Its integral $\int^\infty_0 g(\tau)d\tau\approx 1.19$ characterizes inertial-range timescale \cite{Kaneda86}.}
\label{fig.gtau}
\end{figure}

In principle, Eqs. (\ref{E eq})-(\ref{FD}) may be applicable to arbitrary Reynolds number, while the inertial-range solutions for sufficiently high Reynolds number further simplify the problem to solve. Then the inertial range solution of $\bas Q(\check{k};t,t|\vc{y})$ and $\bas G(\check{k};t,t|\vc{y})$ is obtained from scale-similar solutions of Eqs. (\ref{E eq})-(\ref{FD}), i.e.  
\begin{subequations}
\begin{align}
\bas Q(\check{k};t,t|\vc{y})&=\frac{K_o}{2\pi} \bepsilon^{2/3}(\vc{y},t)\check{k}^{-11/3},
\label{inertial-range Q}\\ 
\bas G(\check{k};t,t|\vc{y})&=g( \bepsilon^{1/3}(\vc{y},t)\check{k}^{2/3}|t-t'|),
\label{inertial-range G}
\end{align}
\end{subequations} 
where $\bepsilon(\vc{y},t)(\equiv 4\pi\nu\int^\infty_0 \check{k}^4\,\bas Q(\check{k};t,t|\vc{y})d\check{k})$ is the dissipation rate of the basic field, $g(\tau)$ is a dimensionless function (see Fig. \ref{fig.gtau}) obtained by solving an integrodifferential equation:
\begin{subequations}
\begin{align*}
&\frac{d}{d \tau}g(\tau)
=-\frac{K_o}{2\pi}\int^\infty_0 dv v^{-2/3} J(v)\int^\tau_0 d\sigma g(v^{2/3}\sigma)\ g(\tau)\\
&\left(\ J(v)=\frac{\pi}{2a^4}\left[(a^2-1)^2 \mathrm{ln} \frac{1+a}{|1-a|}-2a+\frac{10}{3}a^3\right],\ 
a=\frac{2v^2}{1+v^2}\ \right)
\end{align*}
\end{subequations}
with its initial condition $g(0)=1$, and the Kolmogorov constant is numerically solved as $K_o\approx 1.72$ \cite{Kaneda86}. For simplicity of later analysis, we avoid detailed descriptions of energy-containing and dissipation ranges, and instead express them by cutoff wavenumbers $k_c$ and $k_d$. Substituting Eqs. (\ref{inertial-range Q}) and (\ref{inertial-range G}) into Eqs. (\ref{KB})-(\ref{DII}) yields 
\begin{subequations}
\begin{align}
&\bask = \frac{3}{2}K_o \,\bepsilon^{2/3}k_c^{-2/3}\left(1-\bas\mathrm{Re}^{-1/2}\right),\\
&\nu_T = 0.182 K_o\, \bepsilon^{1/3}k_c^{-4/3}\left(1-\bas\mathrm{Re}^{-1}\right),\\
&\gamma_T = 7.35\times 10^{-2} K_o k_c^{-2}\left(1-\bas\mathrm{Re}^{-3/2}\right),\\
&N_I = 2.58\times 10^{-2}K_o k_c^{-2}\left(1-\bas\mathrm{Re}^{-3/2}\right),\\
&N_{I\!I}=0,\\
&N_{I\!\!I\!\!I}=-1.53\times 10^{-2} K_o k_c^{-2}\left(1-\bas\mathrm{Re}^{-3/2}\right),\\
&N_{I\!V}=0,\\
&N_V=6.56\times 10^{-2} K_o k_c^{-2}\left(1-\bas\mathrm{Re}^{-3/2}\right),\\
&D_I=3.34\times 10^{-2} K_o\,\bepsilon^{1/3}k_c^{-10/3}\left(1-\bas\mathrm{Re}^{-5/2}\right),\\
&D_{I\!I}=-1.56\times 10^{-2} K_o\,\bepsilon^{1/3}k_c^{-10/3}\left(1-\bas\mathrm{Re}^{-5/2}\right),\\
&D_{I\!\!I\!\!I}=-2.23\times 10^{-2} K_o\,\bepsilon^{1/3}k_c^{-10/3}\left(1-\bas\mathrm{Re}^{-5/2}\right),
\end{align}
\label{coefficients 0}
\end{subequations}
where $\bas\mathrm{Re} = (k_d/k_c)^{4/3}$ defines a Reynolds number of the basic field. Note that $N_{I\!I}$ and $N_{I\!V}$ vanish for the fluctuation-dissipation relation (\ref{FD}) \cite{MFI}. For fully-developed turbulence we expect $\bas\mathrm{Re}\gg 1$, where $k_d$ less contributes to Eqs. (\ref{coefficients 0}). In contrast, $k_d$ predominantly contributes to dissipation-scale properties. Applying renormalized perturbation analysis of Sec. \ref{FORMULATION} to the total dissipation rate ($\varepsilon$) yields
\begin{equation}
\varepsilon=\bepsilon
+C_\varepsilon\frac{\mathrm{ln}\left(\bas \mathrm{Re}^{-1}\right)}{\bas\mathrm{Re}^{-1}}
\frac{\bask^2}{\bepsilon}\mathsfbi{S}\cdot\mathsfbi{S},
\end{equation}
where $C_\varepsilon=9.33\times 10^{-3}$. The ratio $k_d/k_c$ may increase as turbulence develops, and $\bepsilon$ may converge to $\varepsilon$ for $\bas\mathrm{Re}\to\infty$. In this limit, Eqs. (\ref{coefficients 0}) reads 
\begin{subequations}
\begin{align}
&\bask = \frac{3}{2}K_o \,\varepsilon^{2/3}k_c^{-2/3},\label{bask1}\\
&\nu_T = 0.182 K_o\, \varepsilon^{1/3}k_c^{-4/3},\label{nuT1}\\
&\gamma_T = 7.35\times 10^{-2} K_o k_c^{-2},\\
&N_I = 2.58\times 10^{-2}K_o k_c^{-2},\\
&N_{I\!I}=0,\\
&N_{I\!\!I\!\!I}=-1.53\times 10^{-2} K_o k_c^{-2},\\
&N_{I\!V}=0,\\
&N_V= 6.56\times 10^{-2} K_o k_c^{-2},\\
&D_I=3.34\times 10^{-2} K_o\,\varepsilon^{1/3}k_c^{-10/3},\\
&D_{I\!I}=-1.56\times 10^{-2} K_o\,\varepsilon^{1/3}k_c^{-10/3},\\
&D_{I\!\!I\!\!I}=-2.23\times 10^{-2} K_o\,\varepsilon^{1/3}k_c^{-10/3}.\label{DIII1}
\end{align}
\end{subequations}
Solving Eq. (\ref{bask1}) in terms of $k_c$, Eqs. (\ref{nuT1})-(\ref{DIII1}) can be expressed in terms of $\bask$ and $\varepsilon$. Thence Eq. (\ref{R1}) may be rewritten as  
\begin{equation}
\begin{split}
R^{ij}=&\frac{2}{3}\,\bask g^{ij}
-C_\nu\frac{\bask^2}{\varepsilon} S^{ij}\\
&+C_t\frac{\bask^3}{\varepsilon^2}\left(\frac{\mathfrak{D} S^{ij}}{\mathfrak{D} t} + S^i_a S^{ja}\right)\\
&+C'_s\ \boldsymbol{\mathsf{S\cdot S}}g^{ij}\\
&-C_s\frac{\bask^3}{\varepsilon^2}\ S^i_a S^{ja}\\
&+C_c\frac{\bask^3}{\varepsilon^2} \left(S^i_a \Theta^{ja}+S^j_a \Theta^{ja}\right)\\
&+C_{\scalebox{0.5}{MD1}}\frac{\bask^5}{\varepsilon^3} \left(S^i_a{}^{;ja}+S^j_a{}^{;ia}\right)\\
&-C_{\scalebox{0.5}{MD2}}\frac{\bask^5}{\varepsilon^3} \left(\Theta^{i}{}_{a}{}^{;ja}+\Theta^{j}{}_{a}{}^{;ia}\right)\\
&-C_{\scalebox{0.5}{MD3}}\frac{\bask^5}{\varepsilon^3} \Delta S^{ij},
\end{split}
\label{R2}
\end{equation} 
where 
\begin{subequations}
\begin{align}
&C_\nu=4.69\times 10^{-2},\\
&C_t=7.34\times 10^{-3},\\
&C'_s=2.58\times 10^{-3},\\
&C_s=1.53\times 10^{-3},\\
&C_c= 6.56\times 10^{-3},\\
&C_{\scalebox{0.5}{MD1}}=5.00\times 10^{-4},\\
&C_{\scalebox{0.5}{MD2}}=2.33\times 10^{-4},\\
&C_{\scalebox{0.5}{MD3}}=3.33\times 10^{-4}.
\end{align}
\label{constants}
\end{subequations}

\begin{figure}
\centering
\includegraphics[width=12cm]{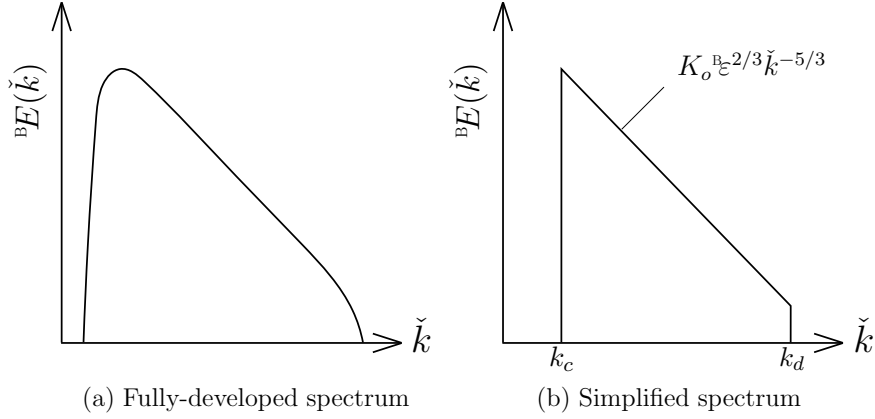}
\caption{(a) For fully developed turbulence, energy spectrum of the basic field may have clear inertial range. (b) For simplicity, we characterize the energy-containing and dissipation ranges by cutoff wavenumbers $k_c$ and $k_d$.}
\label{simplification2}
\end{figure}
Now every transport coefficients incorporates $\bask$ which is not observed from actual turbulence field, and we need to solve $\bask$ in terms of observable variables. Taking the trace of Eq. (\ref{R2}), $\mathscr{K}$ is related with $\bask$: 
\begin{equation}
\mathscr{K}=\bask
+\frac{1}{2}\left(3C'_s-C_s\right)\,\frac{\bask^3}{\varepsilon^2}\mathsfbi{S}\cdot\mathsfbi{S},
\end{equation}
while Eq. (\ref{R2}) may be rewritten as
\begin{equation}
\begin{split}
R^{ij}=&\frac{2}{3}\mathscr{K} g^{ij}
-C_\nu\frac{\bask^2}{\varepsilon} S^{ij}\\
&+C_t\frac{\bask^3}{\varepsilon^2}\left(\frac{\mathfrak{D} S^{ij}}{\mathfrak{D} t} + S^i_a S^{ja}\right)\\
&-C_s\frac{\bask^3}{\varepsilon^2}\left(S^i_a S^{ja}-\frac{1}{3}\mathsfbi{S}\cdot\mathsfbi{S}g^{ij}\right)\\
&+C_c\frac{\bask^3}{\varepsilon^2} \left(S^i_a \Theta^{ja}+S^j_a \Theta^{ja}\right)\\
&+C_{\scalebox{0.5}{MD1}}\frac{\bask^5}{\varepsilon^3} \left(S^i_a{}^{;ja}+S^j_a{}^{;ia}\right)\\
&-C_{\scalebox{0.5}{MD2}}\frac{\bask^5}{\varepsilon^3} \left(\Theta^{i}{}_{a}{}^{;ja}+\Theta^{j}{}_{a}{}^{;ia}\right)\\
&-C_{\scalebox{0.5}{MD3}}\frac{\bask^5}{\varepsilon^3} \Delta S^{ij}.
\end{split}
\label{R3}
\end{equation}
By definition, half the trace of Eq. (\ref{R3}) gives the turbulence energy $\mathscr{K}\equiv R^i_i/2$. By dividing the trace of Eq. (\ref{R3}) by $\mathscr{K}$, we obtain a cubic equation for $\kappa\equiv \bask/\mathscr{K}$:
\begin{equation}
\begin{split}
1=\kappa&
+\frac{1}{2}\left(3C'_s-C_s\right)\,\alpha^2\kappa^3,
\end{split}
\label{kappa eq}
\end{equation}
where $\alpha\,(\equiv \mathscr{K}\norm{\mathsfbi{S}}/\varepsilon)$ is a dimensionless strain rate in a timescale unit $\mathscr{K}/\varepsilon$. Since $3C'_s-C_s > 0$, the right side of Eq. (\ref{kappa eq}) is monotonically increasing. Thus Eq. (\ref{kappa eq}) has only one real solution. Then Eq. (\ref{kappa eq}) can be algebraically solved using Cardano's formula:
\begin{equation}
\kappa=\sqrt[3]{
\left(\bar{\alpha}/2\right)+\sqrt{\left(\bar{\alpha}/2\right)^2+\left(\bar{\alpha}/3\right)^3}
}
+\sqrt[3]{
\left(\bar{\alpha}/2\right)-\sqrt{\left(\bar{\alpha}/2\right)^2+\left(\bar{\alpha}/3\right)^3}
},
\label{kappa}
\end{equation}
where $\bar{\alpha}\equiv 2/(3C'_s-C_s)\alpha^{-2}$ ($\alpha>0$). Now $\kappa(\alpha)$ is a function of the dimensionless strain rate $\alpha$. Obviously $\kappa(\alpha=0)=1$, which is satisfied by Eq. (\ref{kappa}) for the limit $\alpha\to +0$. Using $\kappa(\alpha)$, the Reynolds stress is now given in an explicit form in terms of observable variables $\mathscr{K}$, $\varepsilon$, $\mathsfbi{S}$, and $\mathsfbi{\Theta}$: 
\begin{equation}
\begin{split}
R^{ij}=&\frac{2}{3}\mathscr{K} g^{ij}
-C_\nu\kappa^2(\alpha)\frac{\mathscr{K}^2}{\varepsilon} S^{ij}\\
&+C_t\kappa^3(\alpha)\frac{\mathscr{K}^3}{\varepsilon^2}\left(\frac{\mathfrak{D} S^{ij}}{\mathfrak{D} t} + S^i_a S^{ja}\right)\\
&-C_s\kappa^3(\alpha)\frac{\mathscr{K}^3}{\varepsilon^2}\left(S^i_a S^{ja}-\frac{1}{3}\mathsfbi{S}\cdot\mathsfbi{S}g^{ij}\right)\\
&+C_c\kappa^3(\alpha)\frac{\mathscr{K}^3}{\varepsilon^2} \left(S^i_a \Theta^{ja}+S^j_a \Theta^{ja}\right)\\
&+C_{\scalebox{0.5}{MD1}}\kappa^5(\alpha)\frac{\mathscr{K}^5}{\varepsilon^3} \left(S^i_a{}^{;ja}+S^j_a{}^{;ia}\right)\\
&-C_{\scalebox{0.5}{MD2}}\kappa^5(\alpha)\frac{\mathscr{K}^5}{\varepsilon^3}\left(\Theta^{i}{}_{a}{}^{;ja}+\Theta^{j}{}_{a}{}^{;ia}\right)\\
&-C_{\scalebox{0.5}{MD3}}\kappa^5(\alpha)\frac{\mathscr{K}^5}{\varepsilon^3}\Delta S^{ij}.
\end{split}
\label{R4}
\end{equation} 

\subsection{Rheological nature}
\begin{figure}
\centering
\includegraphics[width=10cm]{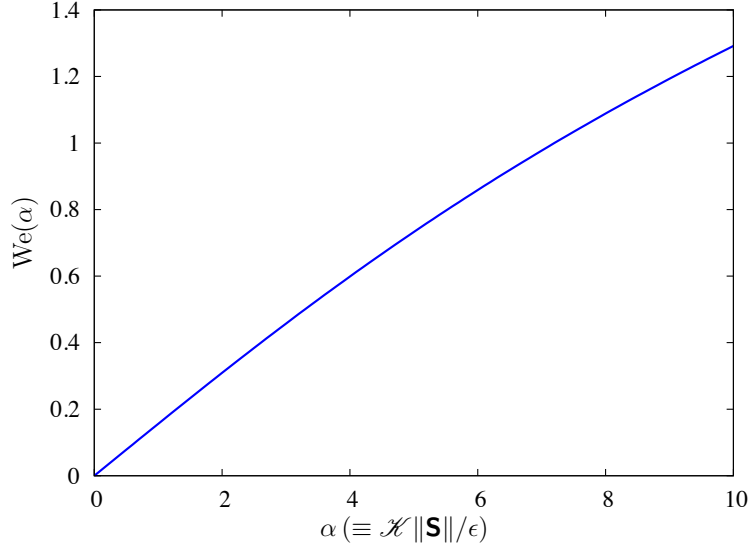}
\caption{Effective Weissenberg number $\mathrm{We}(\alpha)$ shown as a function of the dimensionless strain rate $\alpha$. 
}
\label{weissenberg}
\end{figure}

Rheological nature of turbulence could be understood in a similar manner to dissipation-function formalism in continuum physics. Here we define a \emph{pseudo dissipation function} $\Phi$ by turbulence-energy production rate caused by $\chi^1$- and $\chi^2$-order terms of Eq. (\ref{R4}): 
\begin{equation}
\begin{split}
\Phi&=\frac{1}{4}C_\nu\kappa^2(\alpha)\frac{\mathscr{K}^2}{\varepsilon}\mathsfbi{S}\cdot\mathsfbi{S}
-\frac{1}{8}C_t\kappa^3(\alpha)\frac{\mathscr{K}^3}{\varepsilon^2}\frac{D}{Dt}\mathsfbi{S}\cdot\mathsfbi{S}
+\frac{1}{2}C_s\kappa^3(\alpha)\frac{\mathscr{K}^3}{\varepsilon^2} S^i_j S^j_k S^k_i.
\end{split}
\label{dissipation function}
\end{equation}
The first term comes from the eddy viscosity, giving a positive-semidefinite dissipation. The second term represents the delay response of the viscous dissipation in a similar manner to the viscoelastic fluid. Indeed, one could define a dimensionless factor
\begin{equation}
\mathrm{We}(\alpha)=\frac{\gamma_T}{\nu_T} \norm{\mathsfbi{S}}=\frac{C_t}{C_\nu}\kappa(\alpha)\alpha (\geq 0)
\label{We}
\end{equation}
corresponding to the Weissenberg number. Then the mean flow could be understood as a viscoelastic flow with an effective Weissenberg number $\mathrm{We}(\alpha)$ dependent on dimensionless strain rate $\alpha$. In the present analysis, $\mathrm{We}$ shows a linear dependence on $\alpha$ for $0\leq\alpha\lesssim 10$, which becomes gentle for $\alpha\gtrsim 10$, tending to $\mathrm{We}(\alpha)\propto\alpha^{1/3}$ for $\alpha\to\infty$ (see Fig. \ref{weissenberg}). Reminded that the Wessenberg number quantifies the balance between viscous and elastic timescale, Eq. (\ref{We}) suggests appreciable viscoelasticity of $\mathrm{We}(\alpha\gtrsim 10)=O(1)$. This also implies that the mean flow does not represents prominent elasticity of $\mathrm{We}(\alpha)\gg 1$.

\begin{figure}
\centering
\includegraphics[width=12cm]{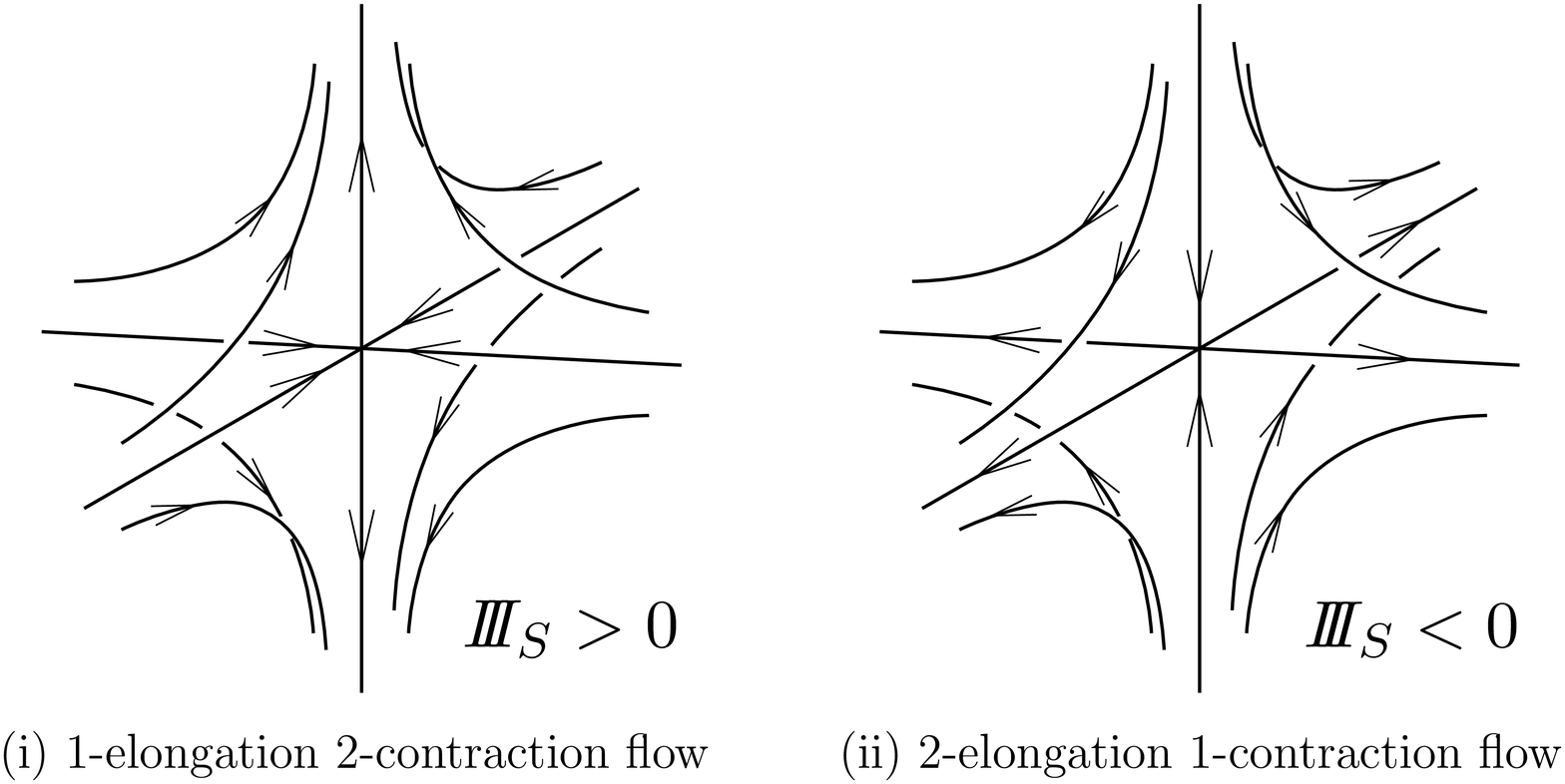}
\caption{The sign of $I\!\!I\!\!I_S$ distinguishes the local straining motion; (i) $I\!\!I\!\!I_S>0$ for 1-elongation 2-contraction flow (ii) while $I\!\!I\!\!I_S<0$ for 2-elongation 1-contraction flow.}
\label{reversible}
\end{figure}

The third term takes both positive and negative values, which may be unusual in continuum physics. In order to understand its anomalous behavior, we utilize the eigen values of $\mathsfbi{S}$. Providing $a,\ b$, and $c$ are the eigen values of $\mathsfbi{S}$, we immediately obtain $S^i_j S^j_k S^k_i=a^3+b^3+c^3$ using principal axes. Now an identity
\begin{equation*}
(a+b+c)^3=a^3+b^3+c^3+3a^2(b+c)+3b^2(c+a)+3c^3(a+b)+6abc
\end{equation*}
subjected to incompressibility condition $a+b+c=0$ yields
\begin{equation*}
0^3=a^3+b^3+c^3+3a^2(-a)+3b^2(-b)+3c^3(-c)+6abc,
\end{equation*}
which reads $S^i_j S^j_k S^k_i=a^3+b^3+c^3=3abc=3I\!\!I\!\!I_S$, where $I\!\!I\!\!I_S$ is the third invariant of the strain rate. In general local flow, following three cases are possible: 
\begin{enumerate}
\item[(i)] $I\!\!I\!\!I_S>0$ when one eigen value is positive and the other two negative, which corresponds to 1-elongation 2-contraction flow (Fig. \ref{reversible}(i) ).
\item[(ii)] $I\!\!I\!\!I_S<0$ when one eigen value is negative and the other two positive, which corresponds to 1-elongation 2-contraction flow (Fig. \ref{reversible}(ii) ).
\item[(iii)] $I\!\!I\!\!I_S=0$ when one eigen value is zero, which corresponds to two-dimensional flow.
\end{enumerate}
Thus, $C_s$-related stress (i) dissipates the mean-flow energy to turbulence energy under 1-elongation 2-contraction flows (ii) converts turbulence energy to mean-flow energy under 2-elongation 1-contraction flows (iii) does not work under two-dimensional flows. Unlike ordinary materials in continuum physics, the function $\Phi$ is no more a positive semidefinite function of strain rate, allowing the back-scatter of the turbulence energy. This is consistent with a known physical picture where fine-scale turbulence is enhanced under rapidly-stretching flows while weakened under contracting flow. TSDIA \cite{Okamoto94} and RNG \cite{RB90} up to second order, in contrast, show opposite trend. 

Unlike $C_\nu$-, $C_t$-, and $C_s$-related stresses, $C_c$-related stress (fifth term on the right side of Eq. (\ref{R4}) ) does not contribute to the turbulence-energy production. This is because $S^\mu_\rho\Theta^{\nu\rho}$ is orthogonal to $S_{\mu\nu}$ in the sense of scalar product; $(\mathsfbi{S}\mathsfbi{\Theta})\cdot \mathsfbi{S}=0$. In general continuum physics, \emph{material-frame-indifference principle} forbids the absolute vorticity to enter the constitutive relation. However, turbulence constitutive relations reflect absolute vorticity due to finite response timescale against rotation, which cannot be understood from known dissipation-function formalism. 

Overall, Eq. (\ref{dissipation function}) is rewritten as
\begin{equation}
\Phi=\frac{1}{4}C_\nu\kappa^2(\alpha)\frac{\mathscr{K}^2}{\varepsilon}\left(1-\frac{\mathrm{We}(\alpha)}{2\norm{\mathsfbi{S}}}\frac{D}{Dt}\right)\norm{\mathsfbi{S}}^2
+\frac{3}{2}C_s\kappa^3(\alpha)\frac{\mathscr{K}^3}{\varepsilon^2} I\!\!I\!\!I_S.
\label{dissipation function2}
\end{equation}
Then, pseudo dissipation function $\Phi$ is no more positive semidefinite but could be negative under suddenly-applied strain or strongly-stretching flow, where turbulence energy could cascade backward to the mean-flow energy.

\subsection{Comparison with channel-turbulence DNS}

\begin{figure}
\includegraphics[width=10cm]{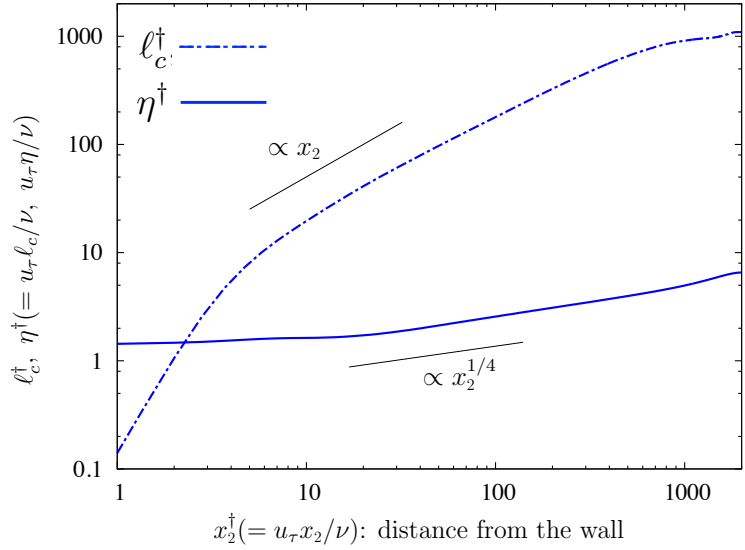}
\caption{Distributions of $\ell_c$ and $\eta$ are shown in the wall unit. We recognize certain separation of scales ($\ell_c/\eta\gtrsim 100$) for outer region $x_{{}_2}^\dagger\gtrsim 200$ $(x_{{}_2}\gtrsim 0.1)$. Power-law behaviors $\ell_c \propto x_{{}_2}$ and $\eta\propto x_{{}_2}^{1/4}$ reflect $\varepsilon\propto x_{{}_2}^{-1}$.}
\label{leta}
\end{figure}

\begin{figure}
\centering
\includegraphics[width=10cm]{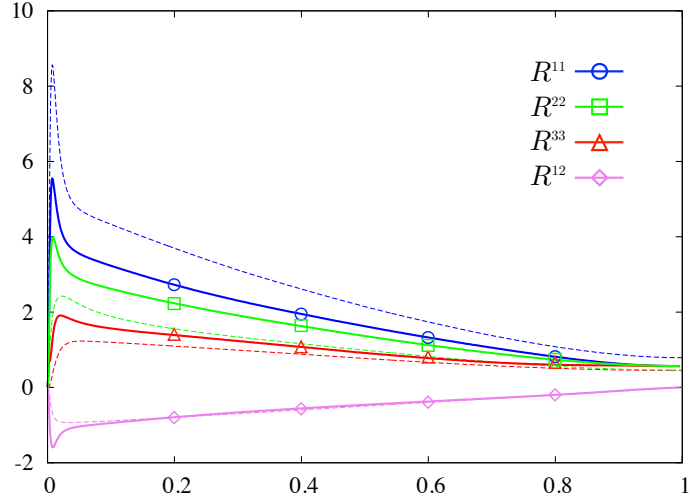}
\caption{Theoretical prediction (\ref{R4}) of the Reynolds stress is plotted by line-point, while dotted lines show DNS data. Velocity and length scales are normalized respectively by the friction velocity and channel half width.}
\label{R comparison}
\end{figure}

\begin{figure}
\centering
\includegraphics[width=16cm]{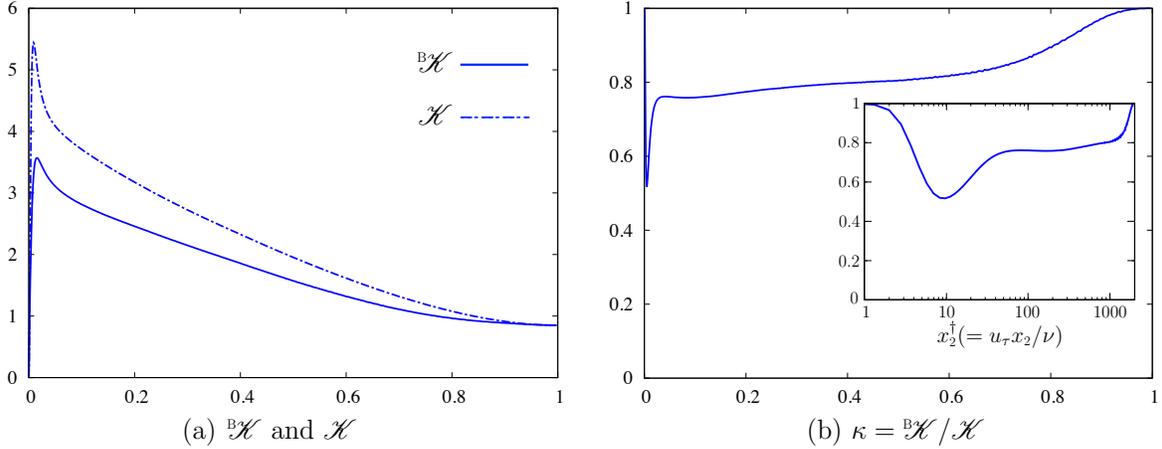}
\caption{(a) ${}^{{}_\mathrm{B}}\!\!\mathscr{K}$, energy of isotropic fluctuation, is appreciably lower than the total energy. (b) The isotropy ratio $\kappa(\alpha)$ is around 0.8 for outer layer of $0.05\lesssim x_{{}_{2}}\lesssim 0.8$. }
\label{Kb comparison}
\end{figure}

\begin{figure}
\centering
\includegraphics[width=9cm]{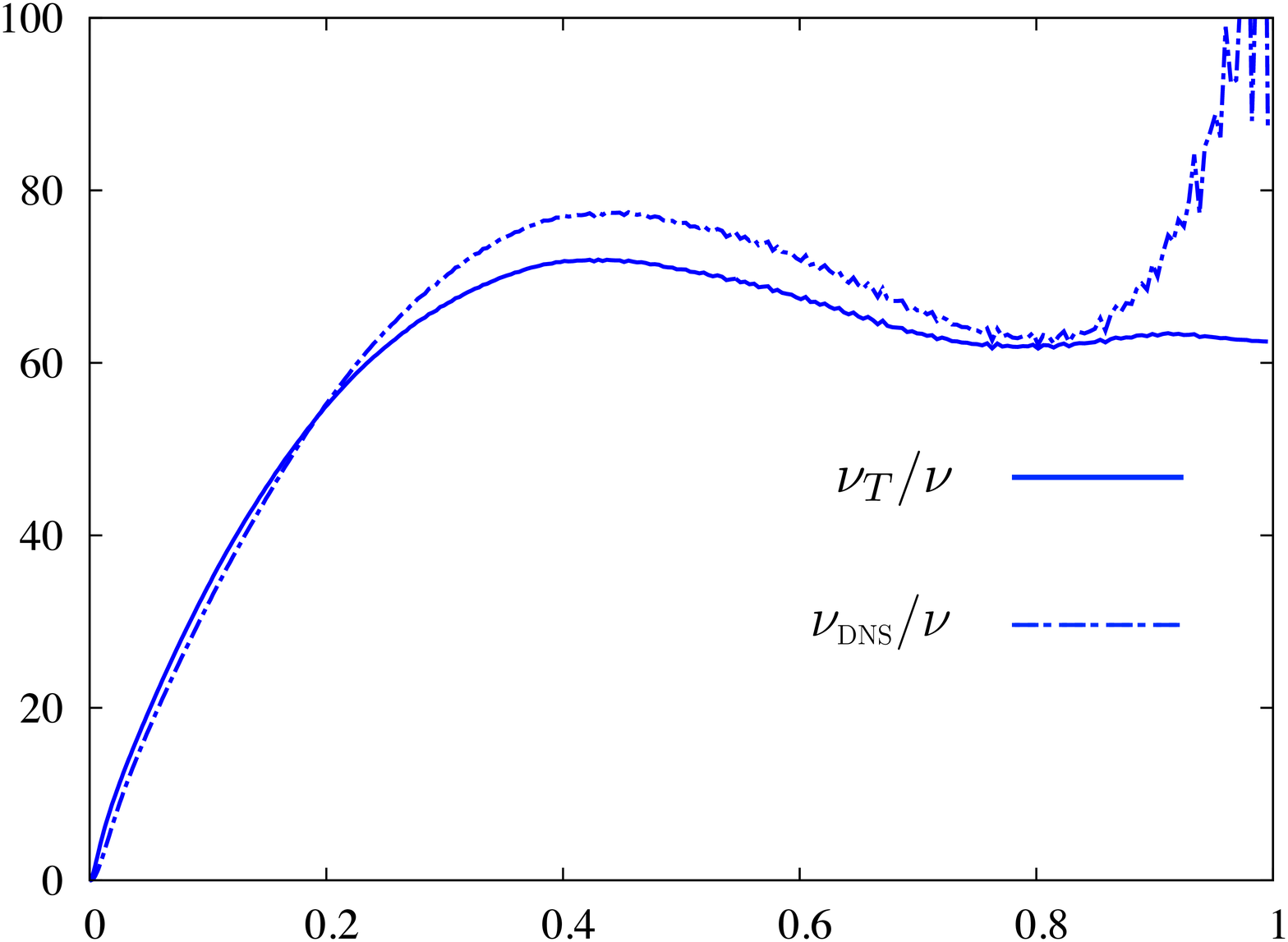}
\caption{Normalized turbulence viscosity $\nu_T/\nu$ (solid line) is compared with an effective turbulence viscosity $\nu_{{}_\mathrm{DNS}}/\nu$ (dotted line). 
}
\label{nuT comparison}
\end{figure}
In order to briefly assess validity of the present analysis, here we perform an a priori test of Eq. (\ref{R4}) using a DNS data of channel turbulent flow of $\mathrm{Re}_\tau=2003$ by Ref. \cite{HJ08}. Orthonormal coordinates $x_{\scalebox{0.5}{1}}$, $x_{\scalebox{0.5}{2}}$, and $x_{\scalebox{0.5}{3}}$ are applied to express the stream-wise, wall-normal, and span-wise directions respectively (here we employ covariant components for convenience in the later discussions). For all the numerical data shown below, the velocity scales are normalized by the friction velocity $u_\tau\equiv\sqrt{\nu d U(x_{\scalebox{0.5}{2}})/d x_{\scalebox{0.5}{2}}}$, while the length scale by channel half width $h$ or by the wall-unit length $\nu/u_\tau$ (let the wall coordinate be $x_{\scalebox{0.5}{2}}^\dagger$ $\equiv u_\tau x_{\scalebox{0.5}{2}}/\nu$).  

Theoretical prediction Eq. (\ref{R4}) is fully based on $k_c\ll k_d$, requiring clear scale separation between the largest- and smallest-eddy sizes. Instead of the wavenumbers $k_c$ and $k_d$, we compare the corresponding length scales $\ell_c\equiv\pi/k_c=\pi(2/3K_o)^{3/2}\ \bask^{3/2}/\varepsilon$ and $\eta\equiv(\nu^3/\varepsilon)^{1/4}$ in Fig. \ref{leta}. We recognize certain separation of scales ($\ell_c/\eta\gtrsim 100$) for $x_{\scalebox{0.6}{2}}^\dagger\gtrsim 200$ $(x_{\scalebox{0.6}{2}}\gtrsim 0.1)$. In the viscous sublayer ($x_{\scalebox{0.6}{2}}^\dagger\lesssim 10$), on the other hand, $\ell_c$ becomes on the same order of $\eta$ (even falls below $\eta$ for $x_{\scalebox{0.6}{2}}^\dagger\lesssim 3$), where $\ell_c$ loses its definition as the largest-eddy scale. 

In Fig. \ref{R comparison}, we compare the Reynolds stress obtained from the DNS and its theoretical prediction from Eq. (\ref{R4}) using the DNS data of $\mathscr{K}$, $\varepsilon$, $\mathsfbi{S}$, and $\mathsfbi{\Theta}$. For diagonal components, an inequality $R^{\scalebox{0.6}{22}}<R^{\scalebox{0.6}{33}}<R^{\scalebox{0.6}{11}}$ are consistently reproduced. The inequality becomes especially important in expressing the secondary stream caused by non-trivial boundary shape in span-wise direction, e.g. one may observe non-trivial span-wise stream in a turbulent flow passing through a straight duct with square cross-section \cite{Speziale87,Vinuesa14}. Although the predicted anisotropy in diagonal components seems somewhat weaker than that in DNS, it could be improved in higher-order analysis, which is left for future studies. 

Regarding $\bask$ as the energy of the isotropic fluctuation, we recognize appreciable difference between isotropic and total fluctuations in Fig. \ref{Kb comparison} (a). It is notable that transport coefficients -- $\nu_T$, $\gamma_T$, $N_I$-$N_V$, and $D_I$-$D_{I\!\!I\!\!I}$ -- are scaled by $\bask$ and $\varepsilon$ in contrast to $\mathscr{K}$-$\varepsilon$ modelings. This is a natural consequence from the fact that the velocity fluctuation $\vc{v}'$ are expressed by linear and non-linear response of isotropic state $\bas \vc{v}$ against non-uniform mean fields. Reminded that $\mathscr{K}$-$\varepsilon$ scaling originates from Kolmogorov's theory of homogeneous-isotropic turbulence, one should employ as a scaling factor the energy representing isotropic fluctuation rather than the total energy $\mathscr{K}$. Indeed, using $\mathscr{K}$ as the expansion basis leads to erroneous behaviors ($R^{\scalebox{0.5}{22}}< 0 < R^{\scalebox{0.5}{33}}\ll R^{\scalebox{0.5}{11}}$, $1\ll-R^{\scalebox{0.5}{12}}$) in the near-wall region of $0\leq x_{\scalebox{0.5}{2}} \lesssim 0.05$. One can further quantify isotropy using $\kappa(\alpha)=\bask/\mathscr{K}$ plotted in Fig. \ref{Kb comparison} (b). In the channel center region, $\kappa=1$ means complete isotropy, while DNS represents weak anisotropic distribution of $R^{\scalebox{0.6}{22}}\lesssim R^{\scalebox{0.6}{33}}<R^{\scalebox{0.6}{11}}$. This may be caused by finite truncation in $\delta$; $\chi^1$, $\chi^2$, and $\delta^2\chi$-order terms yield first and third order derivatives of the mean velocity, which all vanish at channel center. Non-trivial anisotropy appears, for the first time, at $\delta^2\chi^2$-order analysis via second-order derivatives of the mean velocity, which may express the transportation of the anisotropy from boundary layer to the channel center region. $\kappa$ remains around 0.8 in a range $0.05\lesssim x_{\scalebox{0.6}{2}}\lesssim 0.8$, which predicts that the isotropic fluctuation has no less than 80 \% of the total turbulence energy. This may reflect the underestimated anisotropy we found in the diagonal components of the Reynolds stress (see Fig. \ref{R comparison}). In the buffer layer ($5\lesssim{x_{\scalebox{0.6}{2}}}^\dagger \lesssim 30$), strong anisotropy of the Reynolds stress results in a remarkable reduction of $\kappa$. However, as previously remarked, the present analysis cannot be properly applied to the buffer and viscous layers where molecular viscosity plays predominant roles.







In the current analysis, turbulence diffusion of the stream-wise momentum can be consistently quantified by the turbulence viscosity $\nu_T$. In Fig. \ref{nuT comparison},  we compare $\nu_T$ with an effective turbulence viscosity of DNS; $\nu_{\scalebox{0.5}{DNS}}\equiv -R^{{}_{12}}\left/(dV^{{}_1}/dx^{{}_2})\right.$. In a region $0.1\lesssim x_{\scalebox{0.5}{2}}\lesssim 0.8$, $\nu_T$ reasonably predicts $\nu_{\scalebox{0.5}{DNS}}$ with the error being less than 10 \% of $\nu_{\scalebox{0.5}{DNS}}$. Diverging behavior of $\nu_{{}_\mathrm{DNS}}$ in channel center ($x_{\scalebox{0.5}{2}}>0.8$) arises from the velocity gradient $dV^{{}_1}/dx^{{}_2}$ converging to 0.



\section{Discussions}\label{DISCUSSIONS}
In this paper, we have developed a self-consistent closure theory -- TSLRA -- for turbulence constitutive relation on the basis of the two-scale technique and LRA theory. As a natural consequence of double Lagrangian descriptions based on mean and fluctuation velocities, TSLRA offers an organized procedure to describe turbulence correlations in a consistent manner with both Kolmogorov's theory and covariance principle without relying on any empirical parameter. Application of TSLRA to the Reynolds stress successfully yields an extended algebraic expression accompanied by corrections of space-time non-local effect via derivative expansions. The theory reasonably predicts the Reynolds stress obtained from a channel-flow DNS \cite{HJ08}. Although there may be some room for improvement in diagonal components, it is still highly non-trivial that its anisotropic distribution, i.e. $R^{\scalebox{0.5}{22}}<R^{\scalebox{0.5}{33}}<R^{\scalebox{0.5}{11}}$, is consistently reproduced. Note that the inequality $R^{\scalebox{0.5}{22}}<R^{\scalebox{0.5}{33}}$, i.e. imbalance between the wall-normal and span-wise components, plays critically important role in generation of secondary flows under cross-sectional shape of streams \cite{Speziale87,Vinuesa14}. 

The present formulation especially focuses on the general covariance, which should hold to guarantee the physical objectivity of the theory \cite{Ariki15a,Frewer09}. Due to consistency with the general-covariance principle, physics behind turbulence-constitutive relation has been revealed in an objective manner, providing uniform understanding independent of coordinate frames. Thus, for instance, we do not have to carry out separate calculation in the rotating frame to see the frame-rotation effects, unlike TSDIA \cite{YY93,YY93m}. Also TSDIA in the rotating frame results in an apparent contradiction in the physical objectivity of the theory; the mean-vorticity vector $\boldsymbol{\Omega}$ ($\equiv\mathrm{rot}\, \boldsymbol{V}$) in the TS-decomposed equation in TSDIA formalism is accompanied by the frame-rotation vector $\vc{\omega}_p$ in the form of $\boldsymbol{\Omega}/2+2\boldsymbol{\omega}_p$ which eventually breaks the coordinate-transformation rule $\boldsymbol{\Omega}\to\boldsymbol{\Omega}+2\boldsymbol{\omega}_p$ in the resultant Reynolds stress and vortex dynamo. In Refs. \cite{YY93,YY93m}, direct derivations of mean-vorticity effects in the vortex and MHD dynamos are actually avoided but instead are estimated so that the transformation rule $\boldsymbol{\Omega}\to\boldsymbol{\Omega}+2\boldsymbol{\omega}_p$ holds. Regarding the MHD dynamo, Ref. \cite{HS08} has carefully pointed out the breaking of Euclidean invariance and made theoretical improvements by employing the corotational derivative instead of non-covariant Lagrangian derivative (see Eqs. (22) and (24) in Ref. \cite{HS08}).

It should be noted that TSLRA is a natural extension of LRA to inhomogeneous turbulence, and, indeed, TSLRA exactly tends to LRA in the limit of weak inhomogeneity and anisotropy. Due to self-consistent formulation and reliable predictability of LRA for homogeneous isotropic turbulence, TSLRA may well have reasonable performances under moderate anisotropy and inhomogeneity. The problem may occur when either of inhomogeneity or anisotropy become dominant, as frequently observed in real phenomena such as near-wall turbulence. Strictly speaking, finite-order truncation can be validated when the magnitude of expansion basis remains less than unity, otherwise it is fairly difficult to justify such truncation in any perturbation analysis. One should note, however, that tensor expansion of TSLRA is not so divergent, unlike conventional non-linear $\mathscr{K}$-$\varepsilon$ expansions of TSDIA and RNG. To illustrate a rough sketch of this fact, we consider an expansion in terms of anisotropy ($\chi$-expansion). Suppose we perform a $\chi^n$-order expansion of the Reynolds stress ($n\in\mathbb{N}$). Then Eq. (\ref{kappa eq}) may be further generalized to
\begin{equation}
1=\kappa+\frac{1}{2}\left(3C'_s-C_s\right)\,\alpha^2\kappa^3+\cdots+C_\mathcal{A}\mathcal{A}\kappa^{n+1}
\label{kappa eq 2}
\end{equation}
where $C_\mathcal{A}$ is a numerical constant, $\mathcal{A}=O(\alpha^{n},\beta^{n})$ is some invariant of dimensionless strain $\mathscr{K}\mathsfbi{S}/\varepsilon$ and vorticity $\mathscr{K}\mathsfbi{\Theta}/\varepsilon$ ($\beta\equiv\mathscr{K}\norm{\mathsfbi{\Theta}}/\varepsilon$). Note $\kappa=O(\alpha^{-\frac{n}{n+1}},\beta^{-\frac{n}{n+1}})$ for $\alpha,\beta\gg 1$. Then the highest-order term in the series expansion of the Reynolds stress, say $C_{\scalebox{0.5}{(n)}}\bask^{n+1}\mathsfbi{S}^{n}/\varepsilon^{n}$, is evaluated as
\begin{equation}
C_{\scalebox{0.5}{(n)}}\kappa^{n+1}\frac{\mathscr{K}^{n+1}}{\varepsilon^{n}}\mathsfbi{S}^{n}=C_{\scalebox{0.5}{(n)}}\mathscr{K}\times O(1).
\end{equation}
In addition, numerical constant $C_{\scalebox{0.5}{(n)}}$ tends to zero as its order increases. Although it is difficult to write down these constants in explicit forms, a rough estimation suggests $C_{\scalebox{0.5}{(n)}}=O(1/n!)$ for $n\to\infty$; TSLRA expansion could yield a convergent series. Thus one can expect that TSLRA truncated at certain finite order may lead to reasonable results. This is solely because our expansion basis is rather $\bask\norm{\mathsfbi{S}}/\varepsilon$ than $\alpha(=\mathscr{K}\norm{\mathsfbi{S}}/\varepsilon)$. So far, we do not know a priori up to which order in $\chi$ we should expand, and this will be clarified by higher-order analyses in the future works.

Toward more practical turbulence modeling on the Reynolds stress, we should pay special attention to the $\delta$ expansion. As a theory for inhomogeneous turbulent flow, TSLRA systematically incorporates space-time non-local effect. Although such formulation sounds quite reasonable as properly reflecting non-local nature of turbulence, the $\delta$ expansion (see last three terms of Eq. (\ref{R4})) may bring some mathematical complexities when solving mean-flow equation directly combined with the theoretical result; as $\delta$ order becomes higher, the differential order of the mean-flow equation increases, which would remarkably change the stability of solution. Boundary conditions should be also carefully reconsidered, since it might restrict numerical schemes available for given geometries of both flow and boundaries. Then the order of expansion may be hardly decided only from theoretical consideration, but should be carefully selected via simulations on certain variety of cases. In addition, the very near-wall turbulence endowed with strong anisotropy and inhomogeneity may be still a challenging target for the current TSLRA formalism relying on homogeneous isotropic turbulence. At least, the simple perturbative expansion based on $\chi$ and $\delta$ needs some radical improvements such as renormalization of $\chi$ and $\delta$. At this level of closure, dynamical equation of the Reynolds stress itself can be closed, which may have a close relationship with so called Reynolds-stress-transport model.


Finally we shall mention the future prospect of TSLRA toward a complete closure modeling, that requires further analyses on unclosed correlations besides the Reynolds stress. It should be, again, emphasized as a prominent advantage that TSLRA, in principle, is able to close arbitrary unclosed correlations composed of velocity and pressure fluctuations, yielding space-time local expression similar to Eq. (\ref{R4}). In addition, for arbitrary correlations, TSLRA is always consistent with both Kolmogorov's theory \cite{K41a} and general-covariance principle, which are impossible by conventional theories such as TSDIA and RNG. We should remark, however, that TSLRA, as a closure theory based on second-order moments, may well fail to capture essential effects from higher-order moments. A number of DNS and experiments have elucidated significant roles of small-scale structures on large-scale phenomenology, and there is a room for certain improvement in any existing moment-closure theories including LRA and TSLRA. Nevertheless, moment-closure theories provide us valuable knowledge of lower-order moments which often play predominant roles in the mean-field dynamics. Thence the author believes it is still worthwhile to develop the inhomogeneous-turbulence closure following well-established moment-closure theories of homogeneous turbulence. \\\\

\noindent\textbf{Acknowledgement}\\
\noindent The author has greatly benefited from continuous communication with Professor Fujihiro Hamba since the author's Ph.D. student days in The University of Tokyo. This work is supported by JSPS KAKENHI Grant No. (S) 16H06339. The author currently belongs to Department of Aerospace Engineering, Tohoku University, Japan.

\appendix
\def\theequation{\Alph{section}\textperiodcentered\arabic{equation}}
\section{Derivation of Eq. (\ref{expanded eq})}\label{Derive expanded eq}
Here let us derive Eq. (\ref{expanded eq}) from the previous Eqs. (\ref{MLTS eq 2}) and (\ref{MLTS incomp 2}). To begin with, we separate the pressure term and the others in Eq. (\ref{MLTS eq 2}): 
\begin{equation}
\left( ik^\mu +\delta\,\yna^\mu \right)\mathrm{p}'(\vc{k},t|\vc{Y})
+\mathscr{U}^\mu\left[\vc{v}'\right]=0,
\label{p separation}
\end{equation}
where $\mathscr{U}^\mu\left[\vc{v}'\right]$ is a functional of $\vc{v}'$:
\begin{equation}
\begin{split}
\mathscr{U}^\mu\left[\vc{v}'\right]
=&\left( \frac{\partial}{\partial t}+\nu k^2
-2i\delta\nu k_\rho\,\yna^\rho -\delta^2\,\yna^2 \right)
v'^\mu(\vc{k},t|\vc{Y})\\
&+\left( ik_\nu+\delta\,\yna_\nu\right)\cnv{k}{p}{q}v'^\mu(\vc{p},t|\vc{Y})v'^\nu(\vc{q},t|\vc{Y})\\
&+\left( S^\mu_\nu+{\Theta^\mu}_\nu \right)(t|\vc{Y})v'^\nu(\vc{k},t|\vc{Y})
-\delta\,\yna_\nu R^{\mu\nu}(t|\vc{Y})\delta^3_c(\vc{k}|\vc{Y}).
\end{split}
\label{U[v]}
\end{equation}
By multiplying both sides of Eq. (\ref{p separation}) by $k_\mu/ik^2$ we obtain
\begin{equation}
{}^{\scalebox{0.5}{p}}\!\hat{L}\mathrm{p}'(\vc{k},t|\vc{Y})
+\frac{k_\nu}{ik^2}\ \mathscr{U}^\nu\left[\vc{v}'\right]=0,
\end{equation}
where ${}^{\scalebox{0.5}{p}}\!\hat{L}=1-\delta ik_\rho \,\yna^\rho/k^2$, $k=\sqrt{g^{\alpha\beta}k_\alpha k_\beta}$. Using the inverse operator of ${}^{\scalebox{0.5}{p}}\!\hat{L}$, $\mathrm{p}'$ is solved in terms of $\vc{v}'$:
\begin{equation}
\mathrm{p}'(\vc{k},t|\mathbf{Y})
=-{}^{\scalebox{0.5}{p}}\!\hat{L}^{-1}\frac{k_\nu}{ik^2}\ \mathscr{U}^\nu\left[\vc{v}'\right],
\label{solved p}
\end{equation}
By substituting Eq. (\ref{solved p}) into Eq. (\ref{p separation}), we obtain
\begin{equation}
\tilde{P}^\mu_\nu\ \mathscr{U}^\nu\left[\vc{v}'\right]=0,
\label{p elimination}
\end{equation}
where 
\begin{equation}
\tilde{P}^\mu_\nu
=\delta^\mu_\nu-(ik^\mu+\delta\, \yna^\mu)\ {}^{\scalebox{0.5}{p}}\!\hat{L}^{-1}\frac{k_\nu}{ik^2}.
\label{P tilde}
\end{equation}
Then, the dynamical equation (\ref{p elimination}) only contains $\vc{v}'$. Substituting $\vc{v}'=\sol\hat{\mathsfbi{L}}{}^{\scalebox{0.7}{-1}}\ \sol\vc{v}$ into Eq. (\ref{p elimination}) yields the governing equation of $\sol\vc{v}$:
\begin{equation}
\hat{P}^\mu_\nu\ \mathscr{U}^\nu\left[ \sol\hat{\mathsfbi{L}}^{-1}\ \sol\vc{v}\right]=0,
\label{vs eq}
\end{equation}
where an operator $\hat{\mathsfbi{P}}$ is a generalized solenoidal operator:
\begin{equation}
\begin{split}
\hat{P}^\mu_\nu=P^\mu_\rho \tilde{P}^\rho_\nu
=P^\mu_\nu-\delta P^\mu_\rho\,\yna^\rho\,
{}^{\scalebox{0.5}{p}}\!\hat{L}^{-1}\frac{k_\nu}{ik^2},
\end{split}
\end{equation}
which tends to the conventional solenoidal operator $\mathsfbi{P}$ in the homogeneous case ($\delta$-term being 0). Expanding the operators ${}^{\scalebox{0.5}{p}}\!\hat{L}^{-1}$ and ${}^s\hat{\boldsymbol{\mathsf{L}}}^{-1}$; i.e. 
\begin{subequations}
\begin{align}
&{}^{\scalebox{0.5}{p}}\!\hat{L}^{-1}
=1+\delta \frac{ik_\rho}{k^2}\,\yna^\rho+\delta^2 \left(\frac{ik_\rho}{k^2}\,\yna^\rho \right)^2+O(\delta^3),\\
&(\sol\hat{L}^{-1})^\mu_\nu
=\delta^\mu_\nu+\delta \frac{ik^\mu\,\yna_\nu}{k^2} 
+\delta^2 \frac{ik^\mu\,\yna_\rho}{k^2} \frac{ik^\rho\,\yna_\nu}{k^2}
+O(\delta^3),
\end{align}
\end{subequations}
and substituting them into Eq. (\ref{vs eq}) yields the demanded Eq. (\ref{expanded eq}).

\section{$\delta$-effect caused by energy gradient}\label{DELTA}
There is no exact counterpart of $\vc{Y}$-derivative of $\tilde{\vc{v}}$ in conventional homogeneous-turbulence theory, which prevents physical interpretation of $\delta$ effects arising from $\delta\ \vc{\yna}\otimes\tilde{\vc{v}}$. One possible solution to this problem may be to seek for a different expression of $\tilde{\vc{v}}$ with preserving its Gaussian nature. Note that the magnitude of $\tilde{\vc{v}}$ can be naturally evaluated by its energy $\tilde{\mathscr{K}}$ defined by Eq. (\ref{KB}) with replacing $\bas \!Q$ with $\tilde{Q}$:
\begin{equation}
\tilde{\mathscr{K}}(t|\vc{Y})
=\frac{1}{2}\int^\infty_0 4\pi\check{k}^2 d\check{k} \tilde{Q}(\check{k};t,t|\vc{Y}),
\label{Ktilde}
\end{equation}
where $\tilde{Q}$ is an isotropic spectral function of the bare field:
\begin{equation}
\langle \tilde{v}^I(\check{\vc{k}},t|\vc{Y})\tilde{v}^J(\check{\vc{k}}',t|\vc{Y})\rangle
=\frac{1}{2}P^{IJ}(\check{\vc{k}}|\vc{Y})\tilde{Q}(\check{k};t,t|\vc{Y})\delta^3(\check{\vc{k}}+\check{\vc{k}}'|\vc{Y}).
\end{equation}
Then we express $\tilde{\vc{v}}$ by another random-field vector normalized by $\tilde{\mathscr{K}}$:
\begin{equation}
\tilde{v}^\mu(\vc{k},t|\vc{Y})
=\tilde{\mathscr{K}}^{1/2}(t|\vc{Y})\,n^\mu(\vc{k},t|\vc{Y}), 
\end{equation}
where the normalization of $\vc{n}$ is given by
\begin{equation}
\langle n^\mu(\vc{k},t|\vc{Y})n^\nu(\vc{k}',t|\vc{Y})\rangle
=P^{\mu\nu}(\vc{k})N(k;t,t)\delta^3(\vc{k}+\vc{k}')
\end{equation}
which satisfies $\int N(\check{k})d^3\check{k}=1$. So far we only rewrote $\tilde{\vc{v}}$ using $\vc{n}$. Now we set up an assumption $\delta\ \vc{\yna}\otimes\vc{n}=0$, where the inhomogeneity itself is totally attributed to $\tilde{\mathscr{K}}$. This allows us to represent $\delta\ \vc{\yna}\otimes\tilde{\vc{v}}$ in terms of $\tilde{\vc{v}}$:
\begin{equation}
\delta\, \yna_\alpha \tilde{v}^\mu(\vc{k},t|\vc{Y})
=\delta\, \frac{1}{2}\tilde{\mathscr{K}}^{-1}(t|\vc{Y})\yna_\alpha\tilde{\mathscr{K}}(t|\vc{Y})\tilde{v}^\mu(\vc{k},t|\vc{Y}),
\end{equation}
which then yields higher-order derivatives, e.g.
\begin{equation}
\begin{split}
&\delta^2\, \yna_\alpha\,\yna_\beta \tilde{v}^\mu(\vc{k},t|\vc{Y})\\
&=\delta^2\, \left\{\frac{1}{2}\tilde{\mathscr{K}}^{-1}(t|\vc{Y})\yna_\alpha\yna_\beta\tilde{\mathscr{K}}(t|\vc{Y})
-\frac{1}{4}\tilde{\mathscr{K}}^{-2}(t|\vc{Y})\yna_\alpha\tilde{\mathscr{K}}(t|\vc{Y})\yna_\beta\tilde{\mathscr{K}}(t|\vc{Y})\right\}\\
&\ \ \ \ \ \ \ \ \times\tilde{v}^\mu(\vc{k},t|\vc{Y}),
\end{split}
\end{equation}
and so forth. Then $\vc{Y}$-derivative of an arbitrary order can be expressed by $\tilde{\vc{v}}$, and conventional renormalization technique can be applied. For instance,
\begin{equation}
\begin{split}
&\delta^2\langle \yna_\alpha \tilde{v}^\mu(\vc{k},t|\vc{Y})\yna_\beta \tilde{v}^\nu(\vc{k},t'|\vc{Y})\rangle\\
&=\delta^2\, \frac{1}{4}\tilde{\mathscr{K}}^{-2}(t|\vc{Y})\yna_\alpha\tilde{\mathscr{K}}(t|\vc{Y})\yna_\beta\tilde{\mathscr{K}}(t|\vc{Y})\langle\tilde{v}^\mu(\vc{k},t|\vc{Y})\tilde{v}^\mu(\vc{k},t|\vc{Y})\rangle,\\
&\overset{\!\!\scalebox{0.4}{LRA}}{\longrightarrow}
\delta^2\frac{1}{4}\bask^{-2}(t|\vc{Y})\yna_\alpha\bask(t|\vc{Y})\yna_\beta\bask(t|\vc{Y})\langle\bas v^\mu(\vc{k},t|\vc{Y})\bas v^\mu(\vc{k},t|\vc{Y})\rangle.
\end{split}
\end{equation}
Then, instead of Eq. (\ref{R1}), the pure $\delta$ effect yield
\begin{equation}
\begin{split}
R^{ij}=&\frac{2}{3}\,\bask g^{ij}
-\chi \nu_T S^{ij}
+\chi\gamma_t\left(\frac{\mathfrak{D} S^{ij}}{\mathfrak{D} t} + S^i_a S^{ja}\right)\\
&+\chi^2N_I\ \boldsymbol{\mathsf{S\cdot S}}g^{ij}
+\chi^2N_{I\!\!I}\ \boldsymbol{\mathsf{\Theta\cdot \Theta}}g^{ij}\\
&+\chi^2N_{I\!\!I\!\!I}\ S^i_a S^{ja}
+\chi^2N_{I\!V}\ \Theta^i{}_a\Theta^{ia}\\
&+\chi^2N_{V} \left(S^i_a \Theta^{ja}+S^j_a \Theta^{ia}\right)\\
&-\delta^2D'_I\, \bask^{-2}\left(\bask^{;i}\,\bask^{;j}-\frac{3}{4}\bask^{;a}\,\bask_{;a}g^{ij}\right)\\
&-\delta^2D'_{I\!I}\, \bask^{-1}\left(\bask^{;ij}-\frac{1}{3}\Delta\,\bask g^{ij}\right)
\end{split}
\label{R6}
\end{equation}
where 
\begin{equation}
D'_I=\frac{1}{15}\int_0^\infty 4\pi\check{k}^2d\check{k}\ \bas Q(\check{k};t,t)/\check{k}^2,\ \ D'_{I\!I}=\frac{3}{2}D'_I,
\label{D'I-II}
\end{equation}
in the orthonormal frame. It is noticeable that, unlike $\chi$-terms, $\delta^2$-terms can take non-zero values even in the absence of the mean velocity gradient. Inertial-range analysis given by Sec. \ref{INERTIAL RANGE} reduce (\ref{D'I-II}) to
\begin{equation}
D'_I=C_{K1}\frac{\bask^4}{\epsilon^2},\ \ 
D'_{I\!I}=C_{K2}\frac{\bask^4}{\epsilon^2},
\end{equation}
where $C_{K1}=1.93\times 10^{-3}$ and $C_{K1}=2.90\times 10^{-3}$. Unlike in case of Eq. (\ref{R1}), nonzero trace of the $D'_I$-term modifies Eq. (\ref{kappa eq}), which prevents us from algebraically solving $\kappa(\equiv\bask/\mathscr{K})$.


\begin{thebibliography}{99}
\bibitem{TSDIA commutation}
In TSDIA, in contrast, the spectra depend on the mean velocity (see Eqs. (23)-(24) in Ref. \cite{Yoshizawa84}), so the spectra of spatial/time derivative terms are accompanied by spatial/time derivatives of the mean velocity field which apparently violates the covariance principle. In the TSLRA formalism, on the contrary, the dynamical equations do not contain the mean velocity from the very beginning and thus it is totally free from the above difficulties.
\bibitem{renormalization}
The original idea of renormalization of Ref. \cite{Wyld61} is fairly different from the later reverse-expansion technique of Ref. \cite{Kraichnan77,Kaneda81}. Instead of deriving an exact renormalized Green function of Ref. \cite{Wyld61}, Refs. \cite{Kraichnan77,Kaneda81} utilize averaged infinitesimal response function.
\bibitem{intm} The exact (infinite-order) renormalized-perurbation expansion, in principle, offers the rigorous solution by definition. However, actual manipulations demand its lower-order truncation which often fails to describe non-Gaussian nature particularly evident in higher-order correlations.
\bibitem{MFI}
If one chooses ALHDIA, instead of LRA, to close the basic-field properties, fluctuation-dissipation relation (\ref{FD}) does not hold, which results in non-trivial $N_{I\!I}$ and $N_{I\!V}$.
\end{thebibliography}
\end{document}